\documentclass[aps,prl,amsmath,amssymb,floatfix,twocolumn,superscriptaddress]{revtex4-1}
\usepackage{parskip} 
\usepackage{tikz} 
\usepackage{graphicx,ulem}
\usepackage{subfigure}
\usepackage{svg}

\usepackage{amsmath}
\usepackage{bbold}
\usepackage{slashed}
\usepackage{amssymb}
\usepackage{indentfirst}
\usepackage{braket}
\usepackage{array}
\usepackage[colorlinks]{hyperref}
\usepackage{float}
\usepackage[margin=1in]{geometry}

\usepackage{natbib}

\begin{document}
\preprint{APS/123-QED}
\title{Separate surface and bulk topological Anderson localization transitions in disordered axion insulators}
\author{Cormac Grindall}\thanks{These authors contributed equally.}
\affiliation{Department of Physics and Astronomy, Center for Materials Theory, Rutgers University, Piscataway, New Jersey 08854, USA}
\author{Alexander C. Tyner}\thanks{These authors contributed equally.}
\affiliation{Nordita, KTH Royal Institute of Technology and Stockholm University 106 91 Stockholm, Sweden}
\affiliation{Department of Physics, University of Connecticut, Storrs, Connecticut 06269, USA}
\author{Ang-Kun Wu}
\affiliation{Department of Physics and Astronomy, Center for Materials Theory, Rutgers University, Piscataway, New Jersey 08854, USA}
\affiliation{Theoretical Division, T-4, Los Alamos National Laboratory (LANL), Los Alamos, New Mexico 87545, USA}
\author{Taylor L. Hughes}
\affiliation{$^5$Department of Physics and Institute for Condensed Matter Theory, University of Illinois at Urbana-Champaign, Urbana, Illinois 61801, USA}
\author{J. H. Pixley}
\affiliation{Department of Physics and Astronomy, Center for Materials Theory, Rutgers University, Piscataway, New Jersey 08854, USA}
\affiliation{Center for Computational Quantum Physics, Flatiron Institute, 162 5th Avenue, New York, NY 10010}

\date{\today}

\begin{abstract} 

In topological phases of matter for which the bulk and boundary support distinct electronic gaps, there exists the possibility of decoupled mobility gaps in the presence of disorder. This is in analogy with the well-studied problem of realizing separate or concomitant bulk-boundary criticality in conventional Landau theory. Using a three-dimensional axion insulator having clean, gapped surfaces with $e^2/2h$ quantized Hall conductance, we show the bulk and surface mobility gap evolve differently in the presence of disorder. The decoupling of the bulk and surface topology {\color{black}yields a regime that realizes a two-dimensional, unquantized anomalous Hall metal in the Gaussian unitary ensemble (GUE), which shares some spectral and response properties akin to the surface states of a conventional three-dimensional (3D) topological insulator.} The generality of these results as well as extensions to other insulators and superconductors is discussed.

\end{abstract}

\maketitle

\par 
{\it Introduction}:
The resilience of topological insulators (TIs) to disorder\cite{Kane2005,FuKane,bernevig2006quantum,ZhangFT,Xu2006,Ryu2007,Ryu2009,Shindou2009,ryu2010topological,Hassan2010,Schubert2012,Leung2012,Kobayashi2013,Morimoto2014,katsura2016,Chiu2016}, and the utility of their similarly robust gapless surface states, have made these systems of paramount importance for potential technological applications\cite{Budich2020,Wang2015Det,viti2016plasma,konstantatos2018current,lang2017topological,gilbert2021topological,wang2019bi2te3,Tan2016,RevModPhys.80.1083,Sau2010,sarma2015majorana,freedman2003topological}. However, the discovery of axion insulators and higher order topological insulators, which have gapped bulk and {\it{gapped}} boundary excitations, has opened the door to a richer landscape of possible physical phenomena, and has extended our understanding of topological bulk-boundary correspondences\cite{tang2019efficient,zhang2019catalogue,vergniory2019complete,tang2019comprehensive,xu2020high,schindler2018higher,Schindlereaat0346,Benalcazar61,BenalacazarCn,Khalaf2018,calga2019,wieder2020strong,Park2019,xie2021higher,Trif2019,Ghorashi2020,Khalaf2018Si,Khalaf2021,Benalcazar2022}.
\begin{figure}
    \centering
    \includegraphics[width=8cm]{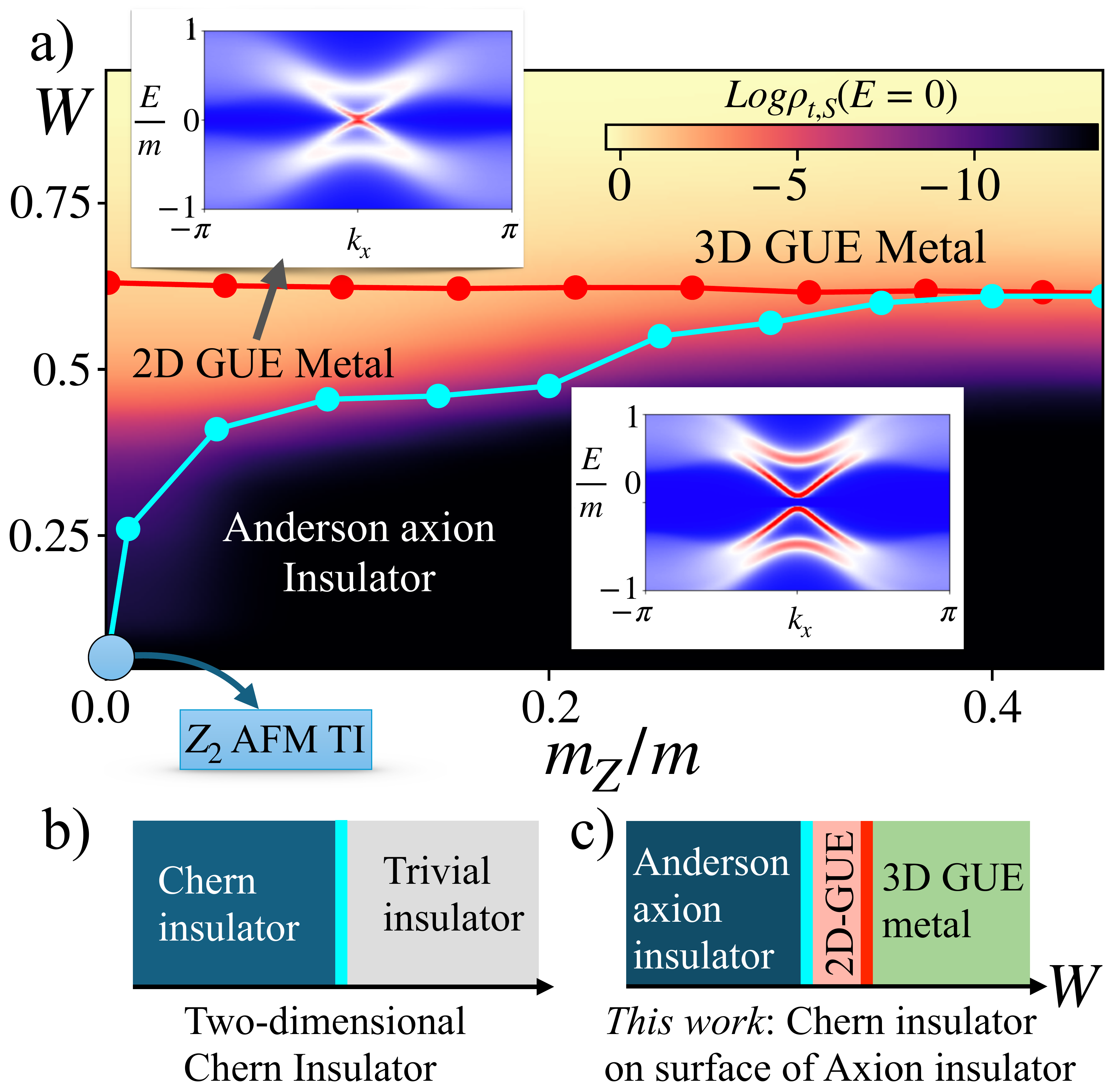}
    \caption{{\bf Phase diagram at the band center of a disordered 3D axion insulator}: a) Surface typical density of states fixing $E/m=0$ as a function of disorder strength $W,$ and staggered Zeeman field $m_{Z}/m$. {Above the teal colored phase boundary, the surface mobility gap is closed and the surface anomalous Hall conductivity is no longer quantized to $\pm e^2/(2h)$. Above the red phase boundary the bulk mobility gap is also closed, coinciding with a change in the bulk topology.} The insets detail results of a disorder averaged surface spectral function for $m_{Z}/m=0.1$, (right) prior to closure of the bulk and surface mobility gap ($W=0.3$) and (left) after closure of the surface mobility gap ($W=0.5$), revealing a massless Dirac cone. Bottom: Schematic of the phase diagram at the band center for (b): a disordered two-dimensional Chern insulator within the GUE class displaying {a closed mobility gap and critical states only at the topological phase transition point (see supplemental\cite{suppl})} and (c): 
    {a half quantized anomalous Hall surface of a 3D axion insulator}
    that first transitions to a two-dimensional metallic phase having an unquantized anomalous Hall conductivity, and then to a three-dimensional metallic phase as disorder is increased.
    }
    \label{fig:fig1}
\end{figure}

The dichotomy between TIs with gapped or gapless surfaces is brought into sharp focus in the context of disordered systems. Conventionally, properties of a disordered TI are controlled by a bulk mobility gap\cite{Xu2006,Hassan2010,Qi2011,Li2020,Hu2021,Wang2020,Wang2021,Wangprb2021,Wang2020,shen2023disorder,Yuliang2023,ding2024disorder,yang2024higher,su2019disorder,agarwala,wang21dis,peng2024,okuma19,yi2024delocalization,nag19,roydis,szabo20,manna22,li2020higher}, %
which can be closed for sufficiently large disorder strengths. 
However, for TIs having gapped boundaries, the separate bulk and surface insulating gaps allow distinct mobility gaps and transitions\cite{Li2020,claes2020,Hu2021,Wang2020,Wang2021,Wangprb2021,Wang2020,Khalaf2021,shen2023disorder,Yuliang2023,ding2024disorder,yang2024higher,su2019disorder,agarwala,wang21dis,peng2024,okuma19,yi2024delocalization,nag19,roydis,szabo20,manna22,li2020higher,Song2021}. This raises the possibility that the boundary mobility gap could close while the bulk mobility gap remains open, hence producing a (de)localization transition on only the surface. If realized, this scenario would represent an Anderson localization analog of the well-studied question of separate versus concomitant bulk-boundary criticality in conventional Landau theory\cite{Fisher1967,diehl1981,diehl1986field,diehl1997theory,metlitski2022boundary}, a topic of recent interest.

In particular, the fate of axion insulators in the presence of disorder is a key open question. Generically, clean axion insulators having broken time-reversal symmetry will have gapped surfaces that exhibit a quantized surface Hall conductance of $e^2/(2h),$ i.e., half the value of the conventional integer quantum Hall effect. In the presence of weak disorder, the surface mobility gap will remain open~\cite{varnava2021controllable}, but {\color{black} stronger disorder may close the surface mobility gap while leaving the bulk mobility gap open. We recall that strong disorder in a two-dimensional (2D) Chern insulator introduces a critical point separating distinct non-trivial and trivial insulating phases that differ by an integer Chern number [shown schematically in Fig.~\ref{fig:fig1}]\cite{Thonhauser2006,prodanets,Prodan2013,moreno2023topological}. In comparison, if a transition were to trivialize the surface of an axion insulator, it would require a direct change in the quantized Hall conductivity by a half-integer. However, only integer valued shifts of $e^2/h$ in the Hall conductivity are expected in non-interacting systems, bringing into question the nature of such a putative surface critical point. 

To this end, we study an axion insulator\cite{Kane2005,FuKane,bernevig2006quantum,ZhangFT,Xu2006,Ryu2007,Ryu2009,Shindou2009,ryu2010topological,Hassan2010,mooreafm,Schubert2012,Leung2012,Kobayashi2013,Morimoto2014,katsura2016,Chiu2016,varnava2021controllable} in the presence of short-ranged quenched disorder. When the surface gap is smaller than the bulk gap in the clean limit, we demonstrate  distinct bulk and surface critical points that 
 are in respective 
agreement
with the 3D Gaussian unitary ensemble (GUE) Anderson localization transition, and the 2D integer quantum Hall (IQH) plateau transition universality classes, respectively. Interestingly, we find that the surface critical point separates an insulating phase with $1/2$-quantized Hall conductance and a 2D GUE metallic phase having a non-quantized surface Hall conductivity 
(but finite mobility gap in the 3D bulk), that is monotonically decreasing in magnitude with increasing disorder strength [see  Fig.~\ref{fig:fig1}(a),(c)]. In this phase, where the surface (bulk) mobility gap is closed (open) [see the region of the phase diagram between the teal and red lines in Fig. \ref{fig:fig1}(a)],
the disorder averaged surface spectral function reveals a characteristic Dirac spectrum [see upper left inset of Fig. \ref{fig:fig1}(a)]. Furthermore, proximity-coupling the surface to an s-wave superconductor reveals properties consistent with the conventional Fu-Kane mechanism\cite{Fu2008} for each disorder sample, i.e., we find Majorana zero modes (MZMs) where vortex lines intersect the proximitized surface.}

\par 
{\it Model: }
We consider the tight-binding model of an antiferromagnetic (AFM) TI on a simple cubic lattice~\cite{varnava2021controllable} in the presence of a quenched, short-range disorder potential: 
\begin{equation}\label{eq:model}
    H=H_{0}+H_{V},
\end{equation}
where $H_{0}=m\sum_{l}c^{\dagger}_{l}\tau^{z}c_{l}+ m_{Z}\sum_{l}(-1)^{l_{z}}c^{\dagger}_{l}\sigma^{z}c_{l}\\+\frac{t}{2}\sum_{\langle l,l '\rangle}c^{\dagger}_{l}\tau^{z}c_{l'}+\frac{-i\lambda}{2}\sum_{\langle l,l'\rangle}c^{\dagger}_{l}\tau^{x}\hat{\mathbf{n}}_{ll'}\cdot \mathbf{\sigma}c_{l'}$, and disorder is introduced as, $H_{V}=\sum_{l}V(l)c^{\dagger}_{l}c_{l}$ where $\tau^{\mu=x,y,z}$ and $\sigma^{\mu=x,y,z}$ are Pauli matrices for the orbital and spin degrees of freedom. Sites on the cubic lattice are labeled by $l$, with $\sum_{\langle l,l'\rangle}$ denoting a sum over nearest neighbors and $\hat{\mathbf{n}}_{ll'}$ the nearest neighbor unit vector. The onsite orbital energies and staggered Zeeman field strength are determined by $m$ and $m_{Z}$ respectively; $t$ dictates the spin-independent hopping strength, and $\lambda$ is the spin-dependent nearest-neighbor hopping strength. The potential $V(l)$ is sampled from a Gaussian distribution with zero mean and variance $W^{2}$, and hence $W$ characterizes the strength of disorder. Throughout this work, $W$ will be expressed in units of $m$. 
\par
The symmetries of this model in the clean limit, $V(l)=0$, are given in the supplementary material of Ref. \cite{varnava2021controllable}, but briefly summarized here. When $m_{Z}=0$, this model reduces to a time-reversal invariant 3D topological insulator 
having a gapped bulk and a quantized axion angle, $\theta=\pi$ for $|t|<|m|<|3t|,$ and $\theta=0$ for $|m|>|3t|$ or $|m|<|t|$. In phases where $\theta=\pi$ the system will have gapless surface states consisting of an odd number of surface Dirac cones\cite{ZhangFT}. In this work we focus on the parameter choice $(m,t,\lambda)=(1.0,-0.5,-0.6)$. For $m_{Z}\neq0$ the staggered Zeeman field gaps the topological surface states on the $z$-surfaces, but quantization of $\theta$ remains, yielding an axion insulator\cite{AxCouple,MagTI}. The insulating $z$-surfaces have a quantized half-integer anomalous Hall effect\cite{axionsurf}. Unless stated otherwise, we focus on the case $m_{Z}/m=0.1$ in the main text. 

\begin{figure}
    \centering
    \includegraphics[width=8cm]{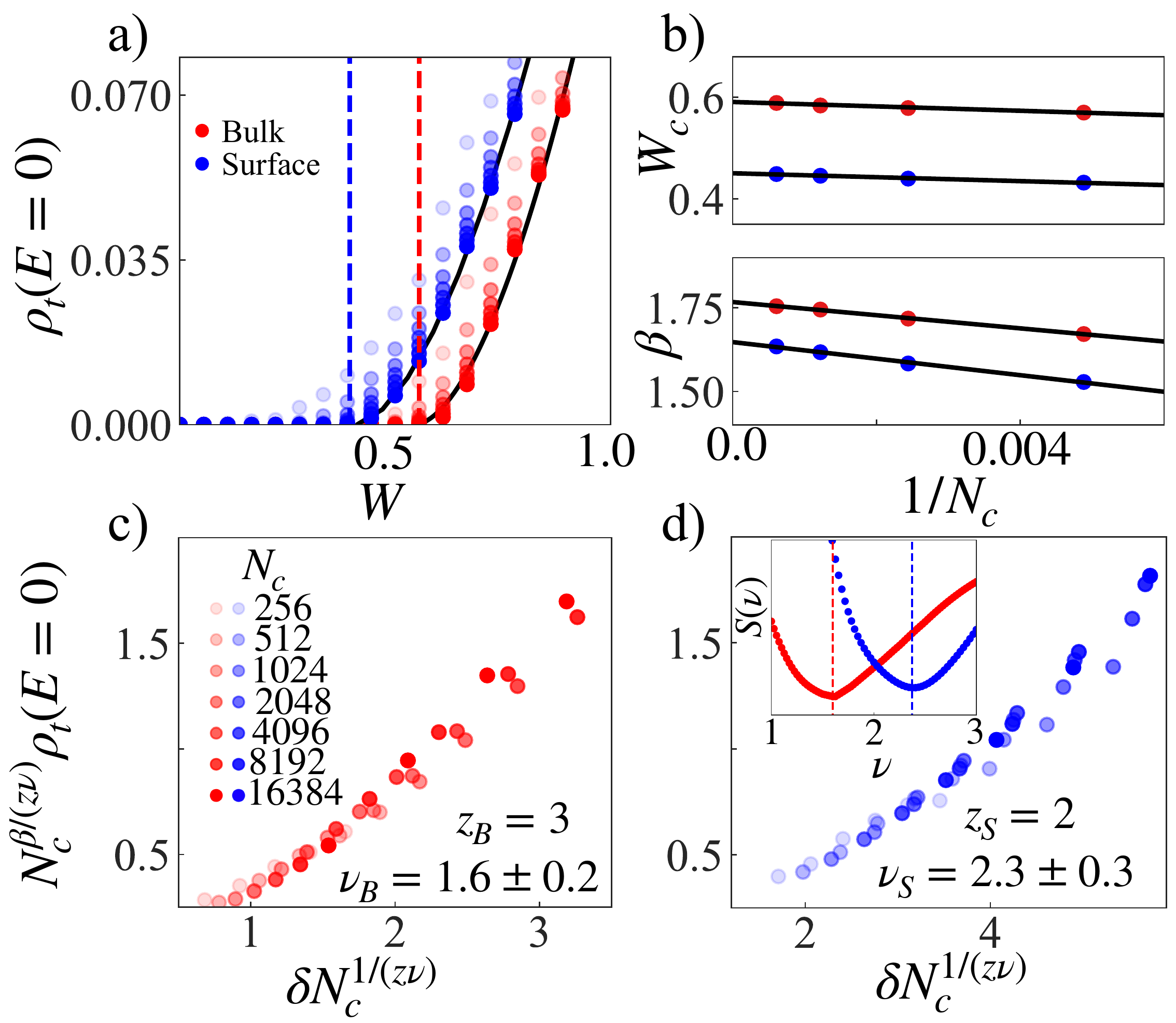}
    \caption{\textbf{Bulk and surface typical density of states:} Fixing $m_{Z}/m = 0.1$ we vary $W$ to pass through each phase of the model shown in Fig.~\ref{fig:fig1}. (a) Red points mark bulk typical density of states at at $E = 0$ and blue points mark surface typical density of states at the band center $E= 0$ as a function of KPM expansion order $N_{c}$ and disorder strength $W$. Black lines are determined by fitting points to $\rho_{t} \sim |W -W_{c,j} |^{\beta_{j}}$ where $j=B,S$ for the bulk and surface respectively. Insulator-to-metal transition points for the surface and bulk are marked by vertical dashed lines. (b) Fitting $W_{c}$ and $\beta$ as a function of $1/N_{c}$ for the surface and bulk to determine each quantity in the limit $N_{c} \rightarrow \infty$. (c) Data collapse of bulk TDOS, fixing $z = 3$ and utilizing values of $\beta$ and $W_{c}$ as determined from (b). The exponent $\nu_{j}$ is determined by the minimum of a similarity function $S(\nu_{j})$ shown in inset of (d). (d) Data collapse for surface TDOS, fixing $z = 2$ and determining $\nu$ using similarity function.}
    \label{fig:tdos}
\end{figure}
\par

{\it Bulk and surface delocalization transitions}:
A phase diagram in the presence of disorder 
and staggered Zeeman field $m_Z$ is constructed by computing the typical density of states (TDOS). 
The TDOS involves computing the geometric mean of the local density of states (defined below) which is not self-averaging, allowing extended and localized states to be distinguished in disordered systems. Identifying where the TDOS becomes finite in the thermodynamic limit is therefore 
a diagnostic of Anderson localization phase boundaries in any dimension\cite{janssen1998statistics,dobrosavljevic2003typical,kpm,Pixley-2015}. 
\par 
We define the TDOS through the local DOS via
\begin{equation}\label{eq:tdos}
\rho_{\mathrm{t}}(E)=\text{exp}\left(\frac{1}{4N_s}\sum_{i=1}^{N_s}\sum_{\sigma=1}^{2}\sum_{\tau=1}^{2}[\ln\rho_{i,\sigma,\tau}(E)]\right),
\end{equation} 
where $[\dots]$ denotes a disorder average,  $N_s$ indicates a small collection 
of random sites to improve the average ($N_s \ll L^3),$ and the local DOS at site $i$ for spin $\sigma$ and orbital $\tau$, 
is defined as, 
$
    \rho_{i,\sigma,\tau}(E)=\sum_{n}|\langle n|i, \sigma, \tau\rangle|^2\delta(E-E_{n})
$
where $|n\rangle$ and $E_n$ denote an exact eigenstate and its eigenenergy, respectively. 
When examining the surface DOS, the sum over $i$ is restricted to values on the surface of the sample. We restrict sampling for the bulk DOS to the half of the system closest to the middle to avoid any boundary effects.  For the computation of the TDOS we average 5000 disorder configurations and use the kernel polynomial method (KPM)\cite{kpm} to expand 
the action of any operator that is dependent on the Hamiltonian on an eigenstate, $|n\rangle$, in terms of Chebyshev polynomials to order $N_c$. This allows us to access a large linear system size of $L=100$. Further details of the KPM method are available in the supplementary material\cite{suppl}. 


\par 
\par
We focus on the TDOS at the band center ($E=0$) since the band structure is symmetric about zero energy in the clean limit, and upon disorder averaging since $V(l)$ is selected from a Gaussian distribution with zero mean (see supplement for data away from the band center\cite{suppl}). As shown in  Fig.~\ref{fig:tdos}(a), the surface TDOS at the band center 
$\rho_t(E=0)$ lifts off from zero well before the bulk does; at large enough disorder $\rho_t(E=0)$ is converged in expansion order and system size (see supplement for data including smaller sizes\cite{suppl}). Near the transitions however, there is a large finite KPM expansion order correction that we take into account using scaling theory of Anderson localization~\cite{evers2008rmp}.
\par
To extract the bulk and surface mobility edges, 
we fit the data in Fig.~\ref{fig:tdos}(a) for $W$ larger than the critical values to a power-law vanishing TDOS 
\begin{equation}
    \rho_{\mathrm{t},j}(E=0)\sim |W-W_{c, j}|^{\beta_{j}}
    \label{eqn:powerlaw}
\end{equation}
(where $j=S, B$ denotes surface and bulk) 
to determine $W_{c,S}$ and $W_{c,B}$ as a function of $N_{c}$; the fits to the largest $N_{c}$ are shown as black lines in Fig.~\ref{fig:tdos}(a) and shown on a log-log scale in the supplement\cite{suppl}.  The values of $W_{c,j}$ are subsequently extrapolated to the limit $N_{c}\rightarrow \infty$ through a linear fit as a function of $1/N_{c}$, shown in the upper panel of Fig.~\ref{fig:tdos}(b). This yields $W_{c,B}=0.59\pm 0.02$ and $W_{c,S}=0.46\pm 0.02$, 
clearly demonstrating the surface localization transition precedes the bulk. We use this procedure to determine the phase diagram in Fig.~\ref{fig:fig1}(a) as we vary the strength of the disorder $W$ and the staggered Zeeman field $m_{Z}/m$. 
\par 
By extrapolating the fits for $\beta_{j}$ (where $j=S,B$) shown in the lower panel of Fig.~\ref{fig:tdos}(b) to the limit $N_{c}\rightarrow \infty$, we find $\beta_{B}=1.76 \pm 0.1$ for the bulk, and $\beta_{S}=1.64 \pm 0.1$ for the surface for $m_{Z}/m=0.1$. The value of $\beta_{j=S,B}$ is explored for additional values of $m_{Z}/m$ in the supplementary material, yielding consistent results\cite{suppl}. Utilizing the values of $\beta$ and $W_{c}$, we fit $\rho_{t}(E=0)$ at the band center to the scaling form in energy~\cite{Pixley-2016B}
\begin{equation}\label{eq:scaling}
   \rho_{t, j}(E=0)\sim N_{c}^{-\beta_{j}/(z_{j}\nu_{j})} f\left(\delta_{j} N_c^{1/(\nu_{j} z_{j})}\right), 
\end{equation}
where $\delta_{j}=|W-W_{c,j}|/W_{c,j}$ and $z_{j}=d_j$ fixing $z_B=3$ and $z_S=2$. The data is then collapsed by optimizing $\nu_{j}$ using a similarity function, $S_j(\nu)$, as shown in the 
inset of Fig.~\ref{fig:tdos}(d). We find $\nu_{B} = 1.6\pm 0.2$ and $\nu_{S}=2.3\pm 0.3$ for the bulk and surface, respectively, as shown in Fig.~\ref{fig:tdos}(c) and (d). 
\par
\par 
\par
\par 
\par 

Based purely on symmetry, the system falls under the GUE since the Zeeman field breaks  time-reversal symmetry. The values of $\nu_j$ determined through the finite size scaling analysis 
are generally in agreement with the expected values for a GUE Anderson localization transition in 3D at $W_{c,B}$ ($\nu_{3D} \approx 1.56$) and the 2D IQH plateau transition at $W_{c,S}$ ($\nu_{2D} \approx 2.25$)\cite{slevin1999,Brndiar2006,slevin2009,slevin2011}. We find that for $W_{c,S}<W<W_{c,B}$ the mobility gap is closed, hence the surface is a 2D GUE metal while the bulk is insulating. To provide further evidence of the GUE metal phase, we performed detailed analysis of the level spacing statistics presented in the supplementary material\cite{suppl}.  

With this evidence in hand, one can ask if the surface critical point differs in any significant ways from the disorder induced topological phase transition 
in conventional Chern insulators as depicted in Fig.~\ref{fig:fig1} (b). First, our estimated value of $\beta_S\approx1.6$ is much larger then what has been found at the IQH transition, {$\beta=0.58$\cite{slevin2009,Amado2011,ujfalusi2015,sbierski2021,Puschmann}.} However, we can seek internal consistency through scaling, as the TDOS exponent can be related to the multifractal nature of the wavefunctions at the localization transition through the expression, $\beta=\nu(\tilde{\alpha}_{0}-d)$\cite{janssen1998statistics}, where $\tilde{\alpha}_{0}$ is a measure of the multifractal spectrum~\cite{evers2008rmp}. The obtained values of $\beta_j$ and $\nu_j$ yield $\tilde{\alpha}^{B}_{0}=4.0 \pm 0.3$  and $\tilde{\alpha}^{S}_{0}=2.6 \pm 0.35$. We find that the obtained value for the bulk is in agreement with the expected value for the 3D unitary class~\cite{ujfalusi2015}, $\tilde{\alpha}^{3D}_{0}\approx 4.0$. {\color{black} The expected value for the 2D IQH transition, $\tilde{\alpha}^{2D}_{0}\approx 2.25$ ~\cite{Evers2008prl}, also lies within the margin of error for the obtained value of $\tilde{\alpha}^{S}_{0}$. Taken together, our analysis of the TDOS data demonstrates the existence of separate bulk and surface transitions and an extended parameter regime in which the disordered two-dimensional surface is a 2D GUE metal.} 
\par
\par 

{\it Surface Spectrum and Topological response: }
While we have demonstrated that the surface mobility gap is closed when $W>W_{c,S}$ we can gain more information by computing the spectral function, defined as,
%
\begin{equation}
    A(\mathbf{k}_{\perp},z;E)=\sum_{n}\sum_{\sigma,\tau}|\langle \mathbf{k}_{\perp},z, \sigma, \tau|n \rangle|^2 \delta(E-E_{n})
\end{equation}
where $\mathbf{k}_{\perp}=(k_{x},k_{y}),$ and $|n\rangle$ and $E_n$ denote the exact eigenstate and eigenenergy respectively. 
We are using a mixed momentum-position basis 
$|\mathbf{k}_{\perp},z, \sigma, \tau\rangle=\frac{1}{\sqrt{L^{2}}}\sum_{x,y}e^{i(k_{x}x+k_{y}y)}c_{x,y,z,\sigma,\tau}^{\dagger}|0\rangle$ for spin component $\sigma,$ orbital component $\tau,$ and we take $z=0$ to be the sample surface. 
%
The surface resolved spectral function is computed using KPM~\cite{PhysRevB.97.235108} and the results are averaged over 100 disorder configurations, considering a system size $L=60$ and $N_{c}=2^{12}$. The results are shown in the inset of Fig. \ref{fig:fig1}(a) which identify clear regions showing characteristic gapped ($W<W_{c,S}$) and gapless ($W_{c,S}<W_{c,B}$) 2D surface Dirac cones.  

\begin{figure}
    \centering
    \includegraphics[width=8cm]{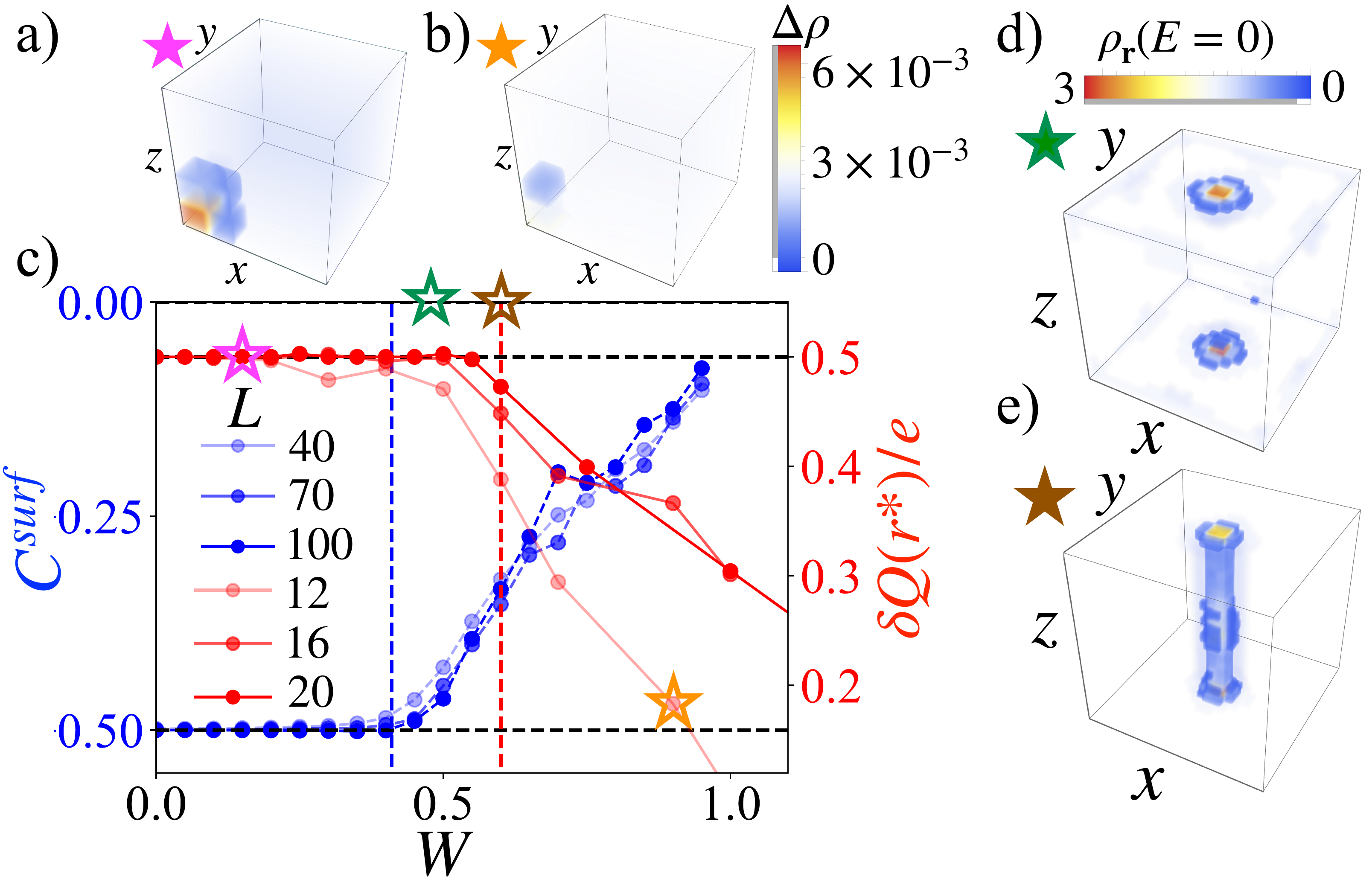}
    \caption{\textbf{Evolution of bulk and surface topology:} (a)-(b) Disorder-averaged, induced charge density within first octant upon monopole insertion. Quantized monopole bound charge {$\delta Q({\bf r})$ } is visible for (a): $W<W_{c,B}$, and absent for (b): $W>W_{c,B}$. (c, red points) Induced charge on monopole at the origin, computed using Eq.~\eqref{eq:chargedens} and exact diagonalization, within a radius $r*=L/4$ about the origin, where $r$ is given in units of the linear system size $L=20$. (c, blue points) Surface Chern marker $C^{\mathrm{surf}}$ on the $\hat{z}$-surface computed using KPM as defined in Eq.~\eqref{eq:chernmarker}, averaged over 5000 disorder configurations fixing $N_c$=16384. The approximate values of disorder for the surface and bulk mobility gap closing are marked by blue and red dashed lines respectively. (d)-(e) Spatially resolved, zero energy density of states of Eq.~\eqref{eq:BDG} fixing $\mu=0,\; m_{Z}/m=0.1$ and $\delta_{0}=0.3$ averaged over fifty disorder configurations for (d) $W_{c,S}<W<W_{c,B}$ and (e) $W=W_{c,B}$.}
    
    \label{fig:Topo}
\end{figure}

While the surface spectral function in the 2D GUE metal regime indicates a gapless Dirac spectrum on average, it is important to understand the response phenomena of the surface and bulk as well. We first consider the topological response of the surface in the presence of disorder. To accomplish this we compute the local Chern marker\cite{cmarker,caio2019topological} as,
\begin{equation}\label{eq:chernmarker}
C_{l}(x,y) = \frac{ 2\pi i}{|\mathcal{A}|} \sum_{s} \bra{s} [\hat{P}_{l}\hat{x}\hat{P}_{l}, \hat{P}_{l}\hat{y}\hat{P}_{l}] \ket{s},    
\end{equation}
where $\mathcal{A}$ is a subsystem constructed of sites on the $z$-layer indexed by $l$, with $|\mathcal{A}|$  its area, and $s$ spans all the degrees of freedom in the subsystem. We compute the local Chern marker via KPM\cite{kpmchern} and we coarse grain across $z$-layers to determine the value on the surface (see  Ref.~\cite{axionsurf} and the supplementary material for more details\cite{suppl}). In the limit of weak disorder, the local Chern marker shows convergence to the quantized half integer value in Fig.~\ref{fig:Topo}. Increasing the disorder strength beyond $W_{c,S}$, we observe a finite value that deviates from a half-integer. This is indicative of the surface being a GUE metal having a non-quantized anomalous Hall effect, {\color{black} even though the average surface spectra indicates a gapless Dirac cone}. 
{\color{black} We confirm the Hall conductance by computing the charge bound to a flux tube passing through the top $z$-surface. We find that such a flux tube binds a charge on the surface in correspondence with the quantized (unquantized) Chern marker for $W<W_{c,S}$ ($W>W_{c,S}$) [see supplementary for details\cite{suppl}].}

By comparison, we can probe the bulk response using the Witten effect\cite{witten1979dyons,yamagishi1983fermion,yamagishi1983fermion2,rosenberg2010witten,zhao2012magnetic,tynermonopole}. That is, we can determine the charge bound to a magnetic monopole inserted in the bulk, which is expected to obey $Q_{M}=\frac{e\theta}{2\pi},$ even in a disordered system.  
To determine the charge bound to the monopole, we compute the charge density at half filling 
\begin{equation}\label{eq:chargedens}
\rho_{q}(\mathbf{r}_i)=-e \sum_{n=0}^{2L^3} |\psi_{n, q}(\mathbf{r}_i)|^2,
\end{equation}
in the presence ($q=M$) and absence ($q=0$) of the monopole. We use a north-pole gauge to simulate the monopole as described in the supplement\cite{suppl}. The induced charge on the monopole is then obtained by integrating
$
\Delta \rho(\mathbf{r}_i)= \rho_M(\mathbf{r}_i) - \rho_0(\mathbf{r}_i)
$
inside a spherical surface of radius $R$ (all lengths are specified relative to the lattice spacing), centered at the monopole. For our discrete system, we explicitly compute 
$
\delta Q(R)=\sum_{|\mathbf{r}_{i}|<R} \Delta \rho(\mathbf{r}_{i})
$ for a system having open boundaries along all directions.
As shown in Fig. \ref{fig:Topo}(a), charge is sharply localized on the monopole and, up to finite size effects, the bound charge is quantized to $Q_M=\tfrac{e}{2}$ all the way to the bulk transition, i.e., even when the surface mobility gap is closed. Hence, we find a robust value $\theta=\pi$ in the bulk for the entire region $W<W_{c,B}$. After the bulk transition there can still be charge density near the location of the monopole, but it is not quantized and depends on the details of the inhomogeneity induced by the disorder. 
\par 
In addition to the electromagnetic response, we can probe the bulk and surface topology by proximity coupling the system to an s-wave superconductor as in the work of Fu and Kane\cite{Fu2008}. In the clean limit, if the proximity-induced superconducting gap dominates over the magnetic surface gap, then vortices that intersect the surface will bind MZMs. Hence, in the regime of disorder where the surface gap has closed but the bulk gap remains open, $W_{c,S}(m_z)<W<W_{c,B}$, we expect that even weak superconductivity should be able to open a surface gap to exhibit the Fu-Kane phenomenology. 
To model this scenario we use a Bogoliubov-de Gennes (BdG) Hamiltonian of the form, 
\begin{equation}\label{eq:BDG}
    H^{BdG}=\begin{bmatrix}
    H-\mu && \Delta_{0}\tau_{0}\sigma_{2} \\
    (\Delta_{0}\tau_{0}\sigma_{2})^{\dagger} && -H^{*}+\mu
    \end{bmatrix},
\end{equation}
$H$ is defined in Eq.~\eqref{eq:model} and $\Delta_0$ is the s-wave pairing amplitude. 
To generate MZMs we can follow the procedure in Refs. \cite{Hosur2011,chiu2011,Rossi2020}, to insert a vortex line along the $z$-axis (see supplementary material\cite{suppl}). The results are shown in Fig. \ref{fig:Topo}(d) which reveal clear localized MZMs. Furthermore, we show that by tuning the chemical potential, $\mu$, a vortex phase transition vanquishing the MZMs is observable, as expected from, e.g., Refs. \onlinecite{Hosur2011,Rossi2020} (see the supplement\cite{suppl}). 
\par 
Interestingly, in  Fig.~\ref{fig:Topo}(e) we show that the vortex phase transition can also be accessed by tuning the disorder strength while keeping the chemical potential fixed. Using the Lanczos algorithm\cite{lanczos1950iteration} at fixed $\mu$ and $m_{z}/m=0.1,$ we average the spatial localization of the zero-energy states for fifty disorder configurations with the results shown in Fig.~\ref{fig:Topo}(d)-(e). For $W_{c,S}<W<W_{c,B}$, we clearly observe localized MZMs, while when $W=W_{c,B}$ the delocalization transition and hybridization across the vortex line is visible. Hence, we find that the Fu-Kane MZMs are apparent for $W>W_{c,S}$ but are removed by the bulk transition when $W>W_{c,B}.$ 



{\it Discussion and Outlook: }
We have shown that introduction of disorder in an axion insulator supporting a gapped topological bulk and gapped topological surface, leads to the decoupling of the surface and bulk mobility gaps. This decoupling gives rise to a novel phase in which the surface hosts a single Dirac cone (on average) within the GUE, {\color{black} a situation that has previously been found at the IQH plateau transition (which occurs at a point, not over an extended parameter range).
We conjecture that this arises because the Chern number cannot change directly by a half-integer, and broken time-reversal symmetry allows for a non-quantized anomalous Hall metal. There could also be a scenario where a critical phase, not a critical point, exists where the multifractal dimensions of the wavefunctions would remain  ``pinned'' to the 2D IQH transition values. However, our data are inconsistent with this scenario (supplement\cite{suppl}) as the surface density of states converges to a finite value in the thermodynamic limit, i.e., it does not vanish in a power-law fashion as it would at the IQH transition.}
\par 
We expect this work to have widespread consequences, as the class of axion insulators includes magnetic topological insulators\cite{rosenberg2010witten,wormhole} beyond the AFM case we investigated. Many higher-order topological insulators (HOTIs) have been shown to fall into this class\cite{Benalcazar61,BenalacazarCn,Schindlereaat0346}, providing a broad range of known systems that can realize the physics described in this work. 

\acknowledgments{{\it Acknowledgments}: We thank Matthew Foster and David Huse for insightful discussions as well as David Vanderbilt and Justin Wilson for discussions and collaborations on related work.
This work is partially supported by NSF Career Grant No.~DMR-1941569 and the Alfred P.~Sloan Foundation through a Sloan Research Fellowship (C.G., A.W., J.H.P.). Part of this work was performed in part at the Aspen Center for Physics, which is supported by the National Science Foundation Grant No.~PHY-2210452 (A.T., J.H.P.)
A.T. acknowledges the hospitality of the Center for Materials Theory at Rutgers University.
NORDITA is supported in part by NordForsk. A portion of this research was done using services provided by the OSG Consortium \cite{osga,osgb,osg,osg2}, which is supported by the National Science Foundation awards No. 2030508 and No. 1836650. The authors acknowledge the Office of Advanced Research Computing (OARC) at Rutgers, The State University of New Jersey for providing access to the Amarel cluster and associated research computing resources that have contributed to the results reported here. A portion of the computations were enabled by resources provided by the National Academic Infrastructure for Supercomputing in Sweden (NAISS), partially funded by the Swedish Research Council through grant agreement no. 2022-06725. A.-K. W. is partially supported by the U.S. Department of Energy (DOE) National Nuclear Security Administration under Contract No. 89233218CNA000001 through the Laboratory Directed Research and Development Program and was performed, in part, at the Center for Integrated Nanotechnologies, an Office of Science User Facility operated for the DOE Office of Science, under user Proposals No. 2018BU0010 and No. 2018BU0083.}

\bibliography{ref.bib}

\end{document}


\renewcommand{\theequation}{S\arabic{equation}}
\renewcommand{\thetable}{S\arabic{table}}
\makeatletter 
\renewcommand{\thefigure}{S\@arabic\c@figure}
\makeatother

\title{Supplementary Material: Separate surface and bulk topological Anderson localization transitions in disordered axion insulators}
\author{Cormac Grindall}\thanks{These authors contributed equally.}
\affiliation{Department of Physics and Astronomy, Center for Materials Theory, Rutgers University, Piscataway, New Jersey 08854, USA}
\author{Alexander C. Tyner}\thanks{These authors contributed equally.}
\affiliation{Nordita, KTH Royal Institute of Technology and Stockholm University 106 91 Stockholm, Sweden}
\affiliation{Department of Physics, University of Connecticut, Storrs, Connecticut 06269, USA}
\author{Ang-Kun Wu}
\affiliation{Department of Physics and Astronomy, Center for Materials Theory, Rutgers University, Piscataway, New Jersey 08854, USA}
\affiliation{Theoretical Division, T-4, Los Alamos National Laboratory (LANL), Los Alamos, New Mexico 87545, USA}
\author{Taylor L. Hughes}
\affiliation{$^5$Department of Physics and Institute for Condensed Matter Theory, University of Illinois at Urbana-Champaign, Urbana, Illinois 61801, USA}
\author{J. H. Pixley}
\affiliation{Department of Physics and Astronomy, Center for Materials Theory, Rutgers University, Piscataway, New Jersey 08854, USA}
\affiliation{Center for Computational Quantum Physics, Flatiron Institute, 162 5th Avenue, New York, NY 10010}

\maketitle
\onecolumngrid
\tableofcontents
\section{Details of clean limit}
\par
\begin{figure*}
    \centering
    \includegraphics[width=16cm]{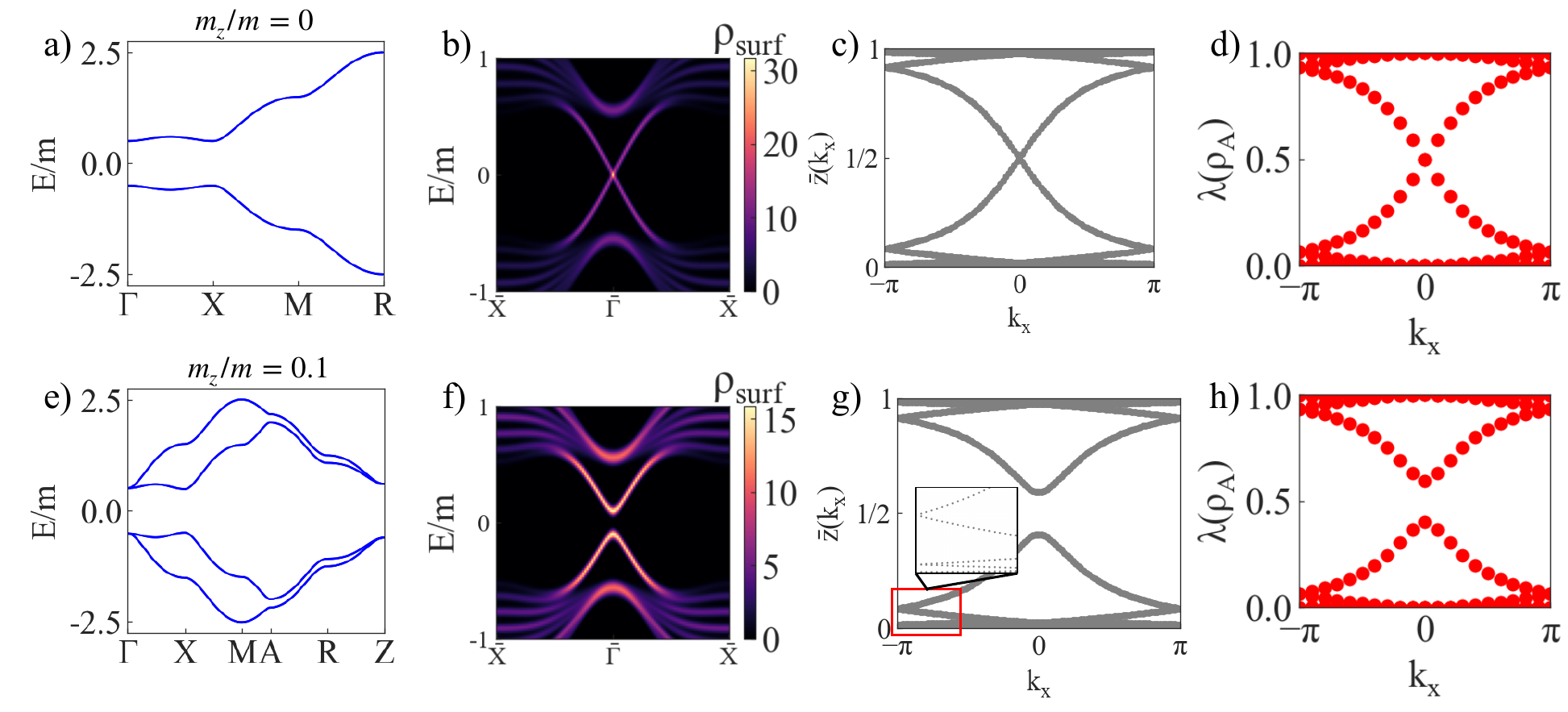}
    \caption{\textbf{Clean properties of the tight-binding model:}(a) Bulk band structure along high-symmetry path for tight-binding model given in Eq.~\eqref{eq:model}, fixing $m_{Z}=0$. (b) Local density of states on the (001) surface considering a slab of size $1\times 1 \times 20$ unit cells and open boundary conditions along the (001) direction fixing $m_{Z}=0$ (c) The gapless Wannier center charge spectra for Eq.~\eqref{eq:model} fixing $m_{Z}=0$ and (d) the gapless entanglement spectra, both represent the bulk-boundary correspondence of the gapless edge states. (e) Bulk band structure along high-symmetry path for tight-binding model given in Eq.~\eqref{eq:model}, fixing $m_{Z}=0.1$. (f) Local density of states on the (001) surface considering a slab of size $1\times 1 \times 20$ unit cells and open boundary conditions along the (001) direction fixing $m_{Z}=0.1$. (g) The gapped Wannier center charge spectra for Eq.~\eqref{eq:model} fixing $m_{Z}=0.1$ and (d) the gapped entanglement spectra. }
    \label{fig:clean}
\end{figure*}

 
The model investigated in the main body is given again here for clarity. In the clean limit, the tight-binding model is of an antiferromagnetic (AFM) topological insulator (TI) proposed in Ref. \cite{varnava2021controllable}, taking the form,
\begin{equation}\label{eq:model}
    H=m\sum_{l}c^{\dagger}_{l}\tau^{z}c_{l}+ m_{Z}\sum_{l}(-1)^{l_{z}}c^{\dagger}_{l}\sigma^{z}c_{l}+\frac{t}{2}\sum_{\langle l,l'\rangle}c^{\dagger}_{l}\tau^{z}c_{l'}+\frac{-i\lambda}{2}\sum_{\langle l,l'\rangle}'c^{\dagger}_{l}\tau^{x}\hat{\mathbf{n}}_{ll'}\cdot \mathbf{\sigma}c_{l'}.
\end{equation}
Sites on the cubic lattice are denoted as $l$. The onsite energies and staggered Zeeman field are determined by $m$ and $m_{Z}$ respectively. Nearest neighbor hoppings are denoted by sums over $\langle l, l' \rangle$; $t$ dictates the spin-independent nearest-neighbor hoppings; $\lambda$ dictates the spin-dependent nearest-neighbor hoppings and $\sigma^{j=x,y,z}(\tau^{j=x,y,z})$ are Pauli matrices that label spin (orbital) degrees of freedom. The nearest-neighbor unit vector is $\hat{\mathbf{n}}_{ll'}$. The finite AFM Zeeman field has the effect of introducing a gap in the $z$ surface spectra. This is reflected in the Wannier center charge (WCC) spectra\cite{Coh2009,yu2011equivalent,Soluyanov2011,alexandradinata2014wilson,Taherinejad2014,Z2pack,bouhon2019wilson,bradlyn2019disconnected}. The WCC spectra is given by the eigenvalues of the Wilson loop, 
\begin{equation}
    W_{z}(k_{x},k_{y})=\mathcal{P}e^{i\oint A_{z}(\mathbf{k})dk_{z}},
\end{equation}
where we define the non-Abelian Berry gauge connection $A_{z}(\mathbf{k})=-i \langle \psi |\partial_{k_{z}}|\psi \rangle$, for the ground state Bloch wavefunction $\psi$. A Wannier obstruction is indicated by the presence of spectral flow in the WCC spectra as a function of varying the transverse momenta. For $m_{Z}=0$ the gapped bulk electronic spectra, gapless surface spectra, and corresponding WCC spectra is seen in \eqref{fig:clean}(a)-(c) respectively. The spectral flow in the WCC spectra constitutes the bulk-boundary correspondence for the gapless edge states. By contrast, when $m_{Z}$ is finite and the surface spectra is gapped, the WCC spectra is correspondingly gapped as seen in Fig. \eqref{fig:clean}(g).  

\begin{figure}
    \centering
    \includegraphics[width=15cm]{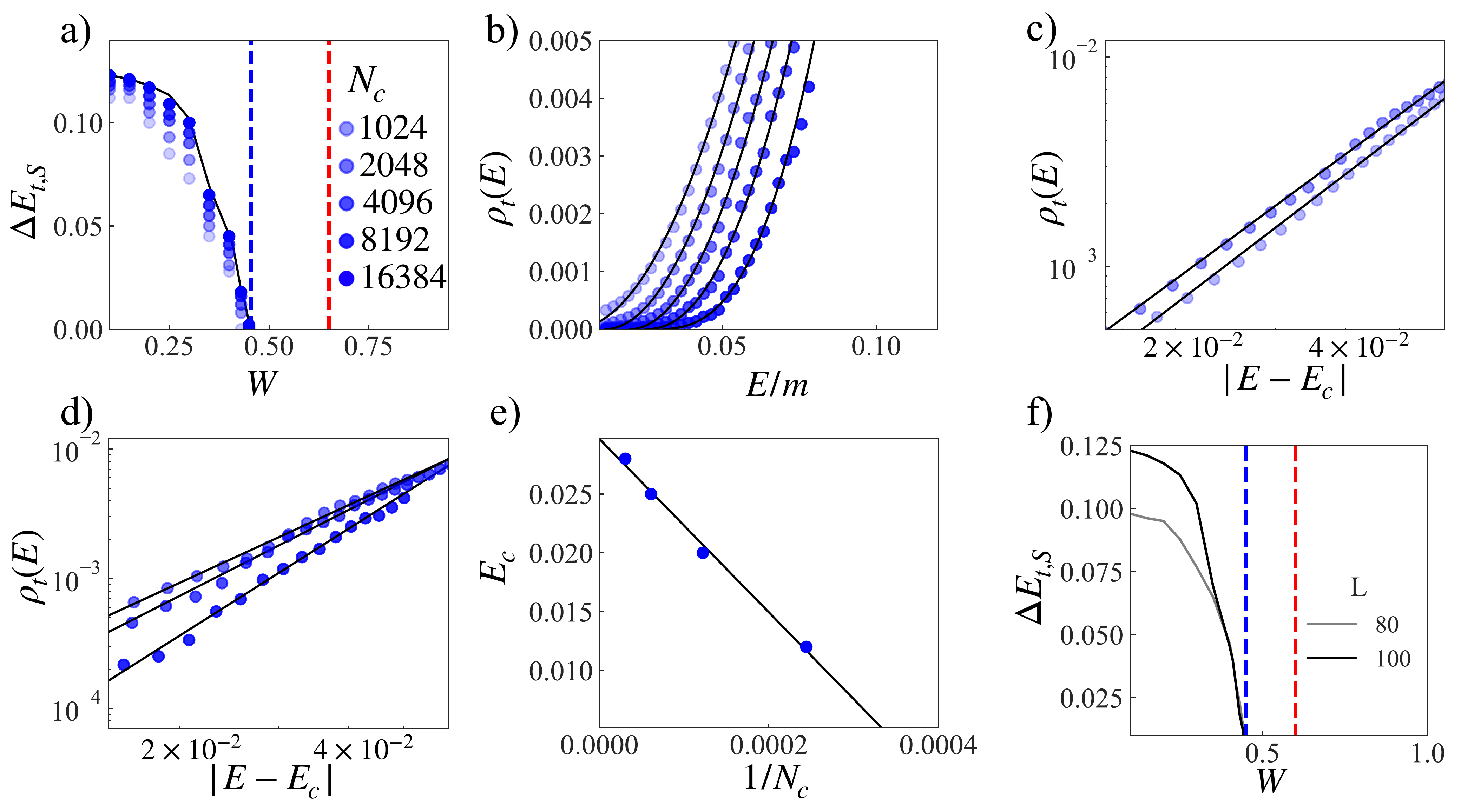}
    \caption{{\bf Tracking the surface typical gap}: (a) The typical gap as a function of the disorder strength, $W$, and the KPM expansion order $N_{c}$ is shown for $L=100$. Blue and red vertical dashed lines mark the surface $(W_{c,S})$ and bulk $(W_{c,B})$ critical points as identified in the main body. Black line details extrapolated value of $\Delta E_{t,S}$ in the limit $N_{c}\rightarrow \infty$. (b) The typical gap is determined by fitting the TDOS to the scaling form $\rho_{t}\sim |E-E_{c}|^{\beta}$. Fits to the TDOS for varying KPM expansion order are shown as black lines, fixing $W=0.42$. To illustrate the quality of the power-law fits, $\rho_{t,S}$ is plotted as a function of $|E-E_{c}|$ on a log-log scale in (c) for $N_{c}=1024, 2048$ and in (d) for $N_{c}=4096, 8192, 16384$. (e) Extracted values of $E_{c}$ as a function of $1/N_{c}$ to determine $E_{c}$ in the limit $N_{c}\rightarrow \infty$. (f) Surface typical gap, $\Delta E_{t,S}$, extrapolated to the $N_{c}\rightarrow \infty$ for two system sizes.     }
    \label{fig:TypGap}
\end{figure}

\par 
We further consider an alternative diagnostic of bulk-boundary correspondence, the entanglement spectrum\cite{prodanets,sondhiets,Qiets1,Qiets2,FidEts,Pollmanets}. To compute the entanglement spectrum we consider creating an imaginary cut which divides the ground state wavefunction into two subsystems, i.e $\psi(\{A_{i}\},\{B_{i}\})$. For a single subsystem we then form the hermitian correlation matrix, $C_{nm}=Tr(\hat{\rho}c^{\dagger}_{n}c_{m})$ , where $c_{n}$ are fermionic operators  with the degrees of freedom, $m,n$, restricted to a single subsystem. The entanglement Hamiltonian, $\hat{H}_A$, can then be obtained from the reduced density matrix, $\rho_{A}$, as
\begin{equation}
    \rho_{A}=e^{-\hat{H}_A}/\mathrm{Tr}(e^{-\hat{H}_A}).
\end{equation}
The eigenvalues of the reduced density matrix, $\lambda(\rho_{A})$, must then fall in the range $[0,1]$. In prior works it has been shown that if a Hamiltonian supports protected edge nodes, the reduced density matrix must also support mid-gap modes, $\lambda(\rho_{A})=1/2$\cite{FidEts,Pollmanets}. This correspondence has been studied both in clean systems as well as in the presence of disorder and shown to be robust\cite{prodanets}. In Fig. \ref{fig:clean}(d) we note the eigenvalues of the density matrix, computed under full periodic boundary conditions are gapless in the limit $m_{Z}=0$ in correspondence with the presence of protected gapless edge states and gapped in Fig. \ref{fig:clean}(h) for $m_{Z}=0.1$ in correspondence with the presence of gapped edge states. In the clean limit the entanglement spectrum thus confers the same information as the WCC spectra. 
\par 

\section{Further analysis of typical density of states}
\par 
\subsection{Surface typical gap}
\par 
In the main body we provide evidence for the existence of a single critical point on the surface, $W_{c,S}$, such that for $W_{c,S}<W<W_{c,B}$ the surface enters a 2D GUE metal phase. To corroborate the evidence presented in the main body for the 2D GUE metal phase we can further track the surface typical gap, $\Delta E_{t,S}$, as a function of disorder strength, $W$. The surface typical gap as a function of the KPM expansion parameter, $N_{c}$ is shown in Fig.~\eqref{fig:TypGap}(a). We observe that the disorder strength at which the typical gap converges to zero is in agreement with the value of $W_{c,S}$ extracted in the main body. The typical gap is determined for each value of disorder strength by fitting the TDOS to the power law, $\rho_{t}\sim |E-E_{c}|^{\beta}$. An example of this is shown for $W=0.42$ in Fig.~\eqref{fig:TypGap}(b). To illustrate the quality of the power-law fits we plot the surface TDOS as a function of $|E-E_{c}|$ on a log-log scale in Fig.~\eqref{fig:TypGap}(c) and Fig.~\eqref{fig:TypGap}(d). A linear fit of $E_{c}$ as a function of $1/N_{c}$ is then performed to extrapolate $E_{c}$ in the $N_{c}\rightarrow \infty$ limit, shown in the inset of Fig.~\eqref{fig:TypGap}(e). The value of $\Delta E_{t,S}$ in the $N_{c}\rightarrow \infty$ is shown by the black line in Fig.~\eqref{fig:TypGap}(a). The value of $\Delta E_{t,S}$ in the limit $N_{c}\rightarrow \infty$ for a system of size $L=80$ utilizing the same procedure for described above for $L=100$ is shown in Fig.~\eqref{fig:TypGap}(f). The results detail that the surface typical gap vanishes at the same value of disorder strength for $L=80$ and $L=100$. This confirms that finite size effects are limited.

\begin{figure}
    \centering
    \includegraphics[width=16cm]{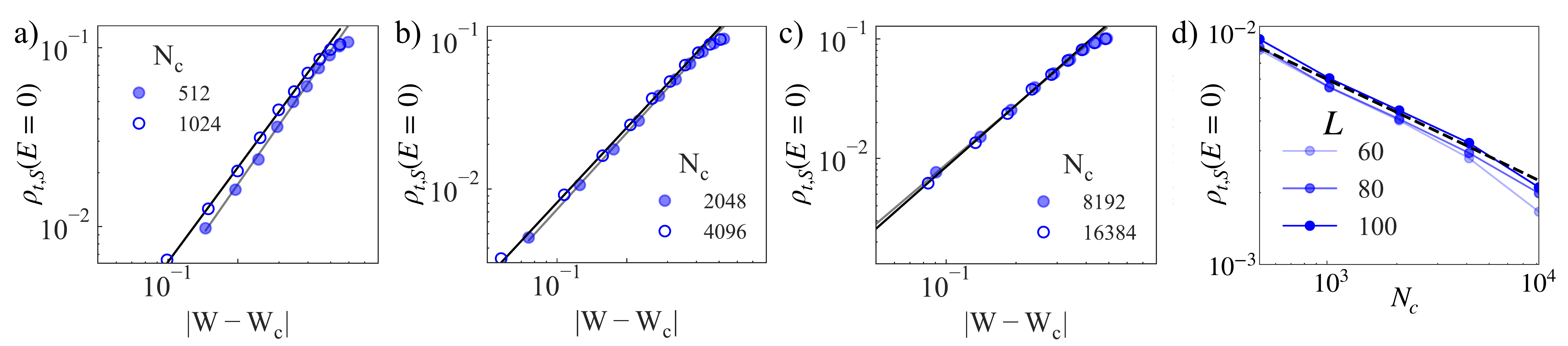}
    \caption{{\bf Scaling at surface critical point}: (a)-(c) Typical density of states at zero energy as a function of $|W-W_{c}|$ on a log-log scale to illustrate quality of the power-law fit, $\rho_{t,S}\sim |W-W_{c}|^{\beta}$ for varying values of the KPM expansion order $N_{c}$. (d) The zero energy typical density of states on the surface at the critical disorder strength, $W_{c,S}$, is shown as a function of $N_{c}$ on a log-log scale for three system sizes. The black-dashed line represents a power-law fit to the data for $L=100$. }
    \label{fig:SurfaceTDOS}
\end{figure}
\par 
\subsection{Scaling near surface critical point}
\par 
In the main body in order to determine the surface critical point we fit the TDOS at zero energy to the power-law form $\rho_{t}\sim |W-W_{c}|^{\beta}$. In order to detail the quality of the power-law fit used to determine $W_{c}$ we further plot the zero energy surface TDOS as a function of $|W-W_{c}|$ on a log-log scale. This is shown in Fig.~\eqref{fig:SurfaceTDOS}(a)-(c) for varying values of the KPM expansion order $N_{c}$. 
\par
In order to perform data collapse and extract the critical exponent, $\nu$, we utilize the scaling form\cite{Pixley-2016B}, 
\begin{equation}\label{eq:scaling}
   \rho_{t}(E=0)\sim N_{c}^{-\beta/(z\nu)} f\left((W-W_{c})N_C^{1/(\nu z)}\right).
\end{equation}
If we choose to focus on the critical point, $W=W_{c,S}$, this relation simplifies to, $ \rho_{t}(E=0)\sim N_{c}^{-\beta/(z\nu)}$. The power-law can further be simplified by noting $\beta/(z\nu)=\alpha_{0}/z-1$. We perform a power law fit of $ \rho_{t}(E=0)$ to check for internal consistency with the critical exponents determined for the surface in the main body. The data, averaging over 1000 disorder configurations for three system sizes, $L$, is shown in Fig.~\eqref{fig:SurfaceTDOS}(d). Fitting the data to a power law form yields $\alpha_{0}=2.8\pm 0.3$. This is within the margin of error for the value determined in the main body. 
\par 

\begin{figure*}
    \centering
    \includegraphics[width=16cm]{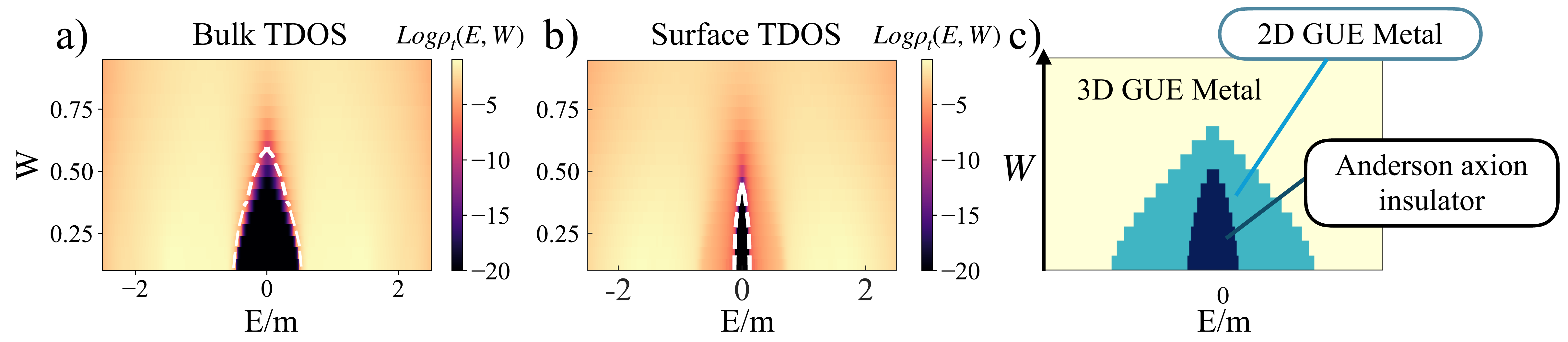}
    \caption{{\bf Finite energy phase diagrams in the surface and bulk}: (a)  The bulk typical density of states (TDOS) defined in Eq. (2) of the main body as a function of disorder strength, $W$ and energy $E$ for a fixed system size of $L=100$ and KPM expansion order $N_C=16384$. Phase boundary is marked by dashed white line computed by extrapolating $W_{c}$ to the limit $N_{c}\rightarrow \infty$. (b) The typical density of states (TDOS) on the surface, as defined in Eq.(2) of the main body,
    restricting the site index to lattice sites on the $z$ surface. This computation is again performed as a function of disorder strength, $W$ and energy $E$ for a fixed system size of $L=100$ and KPM expansion order $N_C=16384$. Phase boundary is marked by dashed white line computed by extrapolating $E_{c}$ to the limit $N_{c}\rightarrow \infty$. (c) Schematic phase diagram as a function of energy and disorder strength fixing $m_{z}/m=0.1$. }
    \label{fig:PhaseMzp1}
\end{figure*}
\subsection{Extended phase diagram for $m_{z}/m=0.1$}
\par 
The scaling and topological analysis presented in the main body was performed fixing the strength of the staggered magnetic field to the value $m_{z}/m=0.1$. Here we present further details of the phase diagram as a function energy and disorder strength, fixing $m_{z}/m=0.1$. The bulk and surface typical density of states is shown in Fig.~\ref{fig:PhaseMzp1}(a)-(b). The white dashed line in each figure marks the phase boundary determined by repeating the scaling analysis detailed in the main body to determine $W_{c,j=B,S}$ in the limit $N_{c}\rightarrow \infty$. Combining the results of Fig.~\ref{fig:PhaseMzp1}(a)-(b) leads to the construction of the schematic phase diagram in Fig.~\ref{fig:PhaseMzp1}(c). 
\par
\begin{figure*}
    \centering
    \includegraphics[width=16cm]{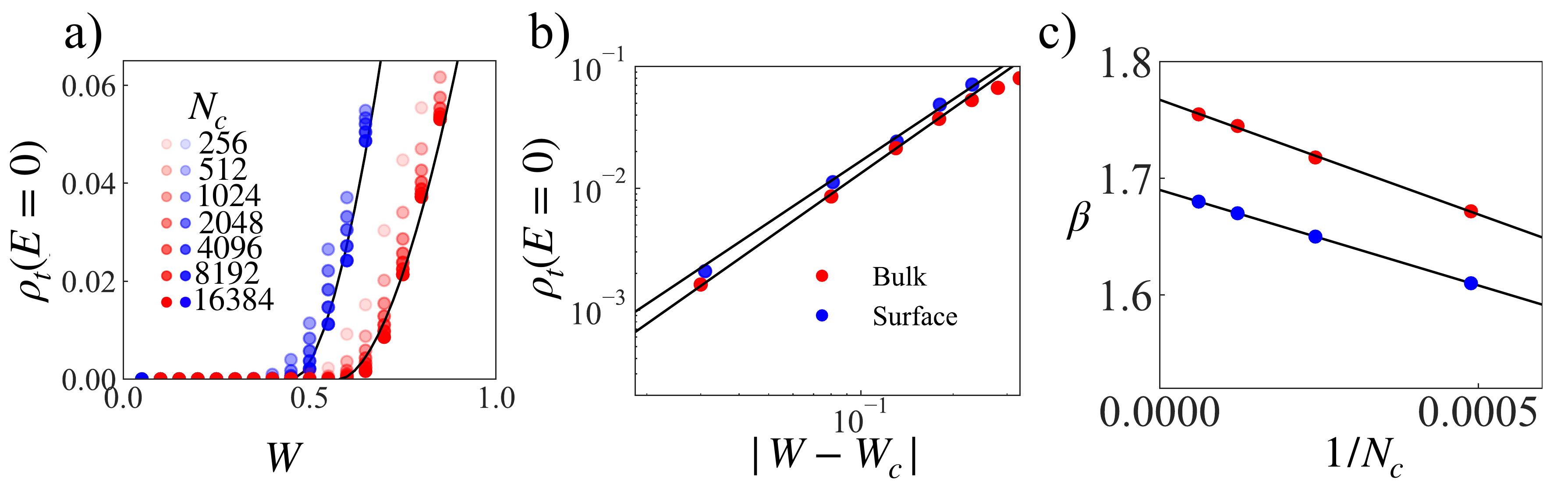}
    \caption{{\bf Scaling at $m_{Z}/m=0.15$}: (a) Red points mark bulk typical density of states at at $E = 0$ and blue points mark surface typical density of states at at $E= 0$ as a function of KPM expansion order $N_{c}$ and disorder strength $W$. Black lines are determined by fitting points to $\rho_{t} \sim |W -W_{c,j} |^{\beta_{j}}$ where $j=B,S$ for the bulk and surface respectively. (b) Fit of TDOS for the largest $N_{c}$ to the power-law vanishing, shown in black in (a), is plotted for the bulk and surface on a log-log scale to determine quality of fit. (c) Fitting $\beta$ as a function of $1/N_{c}$ for the surface and bulk to determine value in the limit $N_{c} \rightarrow \infty$. }
    \label{fig:Mzp15}
\end{figure*}
\subsection{Further scaling details for $m_{z}/m=0.15$}
Throughout this work we have focused primarily on parameter choice, $m_{Z}/m=0.1$. For completeness, here we consider an alternative value, $m_{z}/m=0.15$ and recompute the value of $\beta$ for the surface and bulk, showing that it is consistent with the values derived in main body for $m_{z}/m=0.1$. The TDOS at zero energy as a function of disorder strength is shown for multiple values of $N_{c}$ in Fig.~\ref{fig:Mzp15}(a). We again fit the data in Fig.~\ref{fig:Mzp15}(a) for $W$ larger than the critical values to a power-law vanishing TDOS 
\begin{equation}
    \rho_{\mathrm{t},j}(E=0)\sim |W-W_{c, j}|^{\beta_{j}}
    \label{eqn:powerlaw}
\end{equation}
(where $j=S, B$ denotes surface and bulk) to determine $W_{c,S}$ and $W_{c,B}$ as a function of $N_{c}$; the fits to the largest $N_{c}$ are shown as black lines in Fig.~\ref{fig:Mzp15}(a) and shown on a log-log scale in Fig.~\ref{fig:Mzp15}(b).  
\par
By extrapolating the fits for $\beta_{j}$ (where $j=S,B$) to the limit $N_{c}\rightarrow \infty$ in Fig.~\ref{fig:Mzp15}(c), we find $\beta_{B}=1.75\pm 0.1$ for the bulk, and $\beta_{S}=1.69\pm 0.1$ for the surface consistent with the values obtained for $m_{Z}/m=0.1$.  

\section{Level statistics}

\par 
To provide an additional probe of surface transition and ascertain its diffusive (i.e. random matrix theory like) metallic properties, the level statistics is computed through the adjacent gap ratio, $r_{i}$, and averaged over 500 disorder configurations. We compute the adjacent gap ratio using exact diagonalization as, 
\begin{equation}\label{eq:LevelSpace}
    r_{i}=\frac{\text{min}(\delta_{i},\delta_{i+1})}{\text{max}(\delta_{i},\delta_{i+1})},
\end{equation}
where $\delta_{i}=E_{i+1}-E_{i}$ is the difference between neighboring, distinct eigenvalues.
In the following, the level statistics is computed as a function of energy that we average over 500 disorder configurations for linear system sizes $L=14,16,18,20$ that have periodic boundary conditions in $x$ and $y$ with open boundary conditions along $z$.
\par
\begin{figure}
    \centering
    \includegraphics[width=16cm]{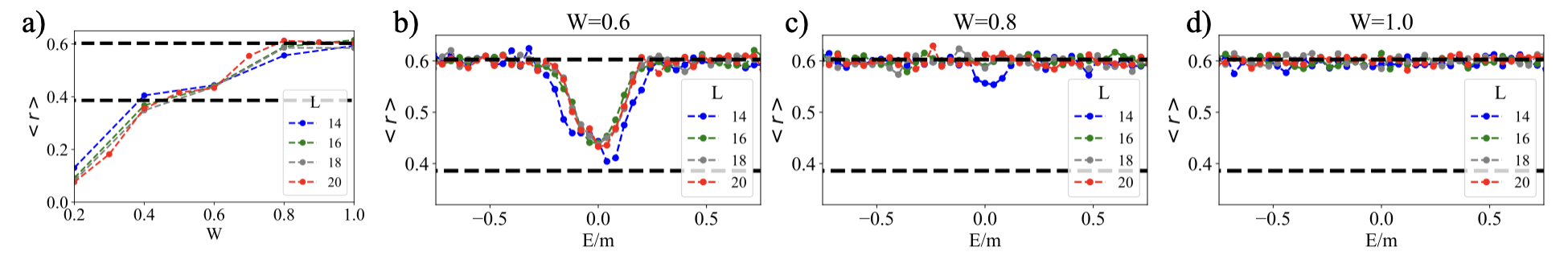}
    \caption{{\bf Level Statistics}: (a) Adjacent gap ratio as defined in Eq.~\eqref{eq:LevelSpace} for $E=0$ as a function of disorder strength, $W$, averaging over 500 disorder configurations. For increasing disorder strength the results obey the expected result for a GUE, $\langle r \rangle \approx 0.602$ (marked with a black dashed line), while at weak disorder the results align with the expected result for a Poisson distribution, $\langle r \rangle = 2\log 2-1 \approx 0.39$. (b)-(c) Adjacent gap ratio for fixed disorder strength (b) $W=0.6$, (c) $W=0.8$ and (d) $W=1.0$ as a function of $E/m$. As disorder strength is increased the adjacent gap ratio near $E=0$ is again shown to transition from the expected result for a Poisson distribution to that of a GUE.   }
    \label{fig:levelstats}
\end{figure}

 
When $m_{z}=0$, this model falls in the three-dimensional symplectic class (in the Cartan classification, class AII)  due to the spin-orbit coupling. However, a non-zero Zeeman field places the system in the three-dimensional unitary class. Upon introduction of disorder the metallic phase can be described by random matrix theory with level statistics that satisfy the Gaussian unitary ensemble (GUE)\cite{levelstats}. For extended states we therefore expect that $\langle r \rangle \approx0.602$, satisfying the GUE while localized states follow the Poisson statistics with  $\langle r \rangle \approx 0.386$. We find this to be the case in Fig. \eqref{fig:levelstats} however we note that these results are impacted by finite size effects as we are unable to access the large system sizes accessible via kernel polynomial method. 
\par
\section{Kernel polynomial method for computation of typical density of states and local Chern marker}
In order to correctly ascertain the localization properties it is essential to reach large system sizes. Therefore TDOS is computed using the kernel polynomial method (KPM)\cite{kpm}, this method expands the quantity of interest in terms of Chebyshev polynomials to an order $N_C$ and is able to reach large system sizes by utilizing efficient sparse matrix-vector multiplication. The system size is sufficiently large such that  the dominant ``finite size'' effect comes from the KPM expansion order denoted $N_C$~\cite{Pixley-2016B}.
\par
A benefit of the local Chern marker is that it can be computed via KPM making large system sizes accessible\cite{cmarker,kpmchern}. As shown in Fig.~\ref{fig:CMLayers}(a) for a system of size $L=100$ averaging over 5000 disorder configurations, bulk layers yield an oscillatory Chern marker. To isolate the surface Chern number, we utilize the sliding widow average over $n$ layers proposed by Ref.~\cite{axionsurf} such that, $C=\sum_{l=0}^{n-1}(C_{l+1}+C_{l})/2$, where $C_{l}$ is the Chern number of layer $l$ computed via Eq.~(8) in the main body. We find convergence for $n=8$. It is important to examine the Chern marker in this layer resolved manner as it establishes that when the surface mobility gap closes for the outermost layer, layers deeper in the bulk which remain insulating do not develop a quantized Chern number.

\section{Details of monopole insertion}
As stated in the main body, the bulk topology of the model we study is characterized by a quantized axion angle, $\theta=\pi$. We determine the axion angle in the presence of disorder via the Witten effect. The Witten effect states that a magnetic monopole inserted into an insulator admitting axion angle $\theta$ binds charge of magnitude, $\frac{e\theta}{2\pi}$\cite{witten1979dyons,yamagishi1983fermion,yamagishi1983fermion2}. For a quantized axion angle, $\theta=\pi$, the monopole thus binds a half-integer electric charge, taking the form of a dyon. In previous works, this has been shown to be robust to the presence of disorder which leaves the bulk mobility gap intact\cite{rosenberg2010witten,zhao2012magnetic,tynermonopole}.  
\par
To determine the charge bound to the monopole, we compute the normalized eigenstates $(\psi_{n, q}(\mathbf{r}_i))$ and energy eigenvalues $(\epsilon_{n,q})$ twice. Once in the presence of the monopole, $q=M$, and once with the monopole absent, $q=0$. The presence of a monopole is simulated using the singular, north-pole gauge
\begin{equation}\label{eq:gauge}
\mathbf{A}(\mathbf{r}_i)= \frac{g}{r_i} \cot \frac{\theta_i}{2} \; \hat{\phi}_i=g \; \frac{-y_i \hat{x} + x_i \hat{y}}{r_i(r_i+z_i)}, 
\end{equation}
where $i$ index lattice sites and we fix $g=1$ to specifically consider the case of a unit-strength monopole. The charge density in the presence and absence of the monopole is then calculated as, 
\begin{equation}\label{eq:chargedens}
\rho_{q}(\mathbf{r}_i)=-e \sum_{n=0}^{2L^3} |\psi_{n, q}(\mathbf{r}_i)|^2 .
\end{equation}
The induced charge density on the monopole is then obtained by integrating
\begin{equation}
\Delta \rho(\mathbf{r}_i)= \rho_M(\mathbf{r}_i) - \rho_0(\mathbf{r}_i),
\end{equation}
inside a spherical Gaussian surface of radius $R$, centered at the monopole. This is done discretely as 
\begin{equation}\label{eq:deltaq}
\delta Q(R)=\sum_{|\mathbf{r}_{i}|<R} \Delta \rho(\mathbf{r}_{i}).
\end{equation} 
All lengths are measured in units of lattice spacing and open boundary conditions are imposed along all directions.

\section{Surface topology via flux-tube insertion}
\par 

\begin{figure}
    \centering
    \includegraphics[width=15cm]{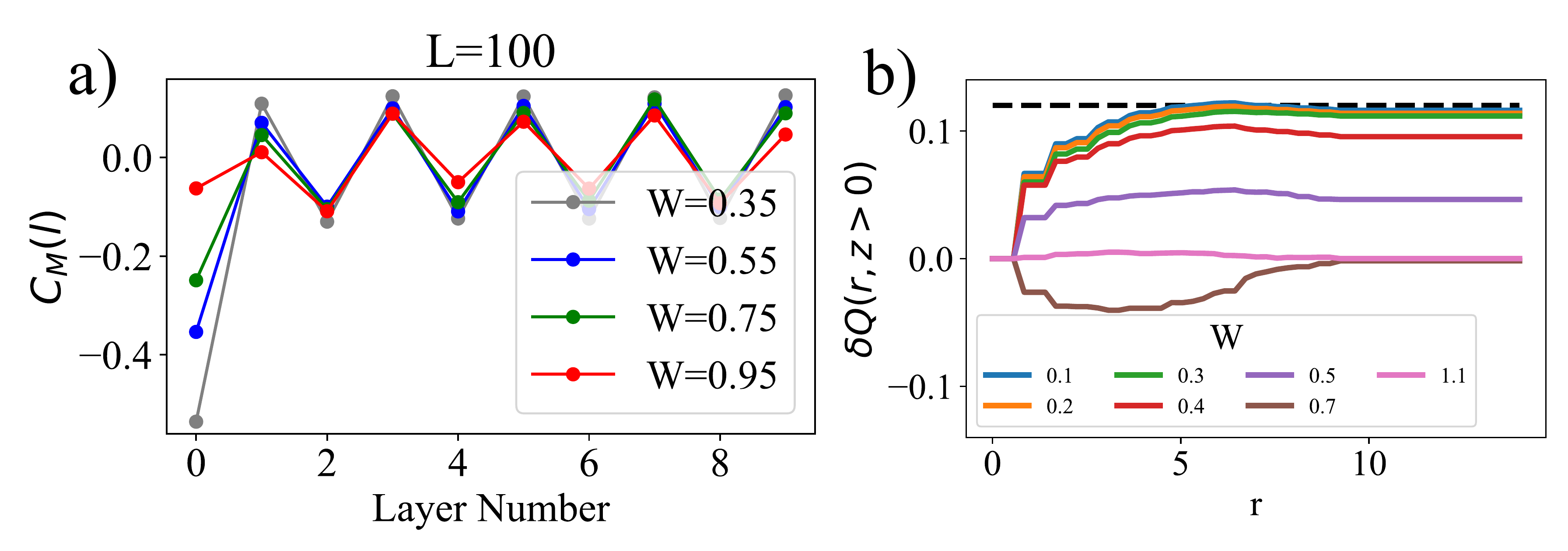}
    \caption{{\bf Surface and Bulk topology}: (a) Disorder averaged chern marker for first 10 $z$-layers as a function of disorder strength for a system of size $L=100$. (b) Induced charge within a cylinder of radius $r$ and $z>0$, upon insertion of flux tube along the $z$ direction of strength $\phi/\phi_{0}=1/4$ for a system of size $L=14$, averaging over 50 disorder configurations. Black dashed line marks expected value of induced charge in the clean limit. }
    \label{fig:CMLayers}
\end{figure}
\par 

\begin{figure*}
    \centering
    \includegraphics[width=14cm]{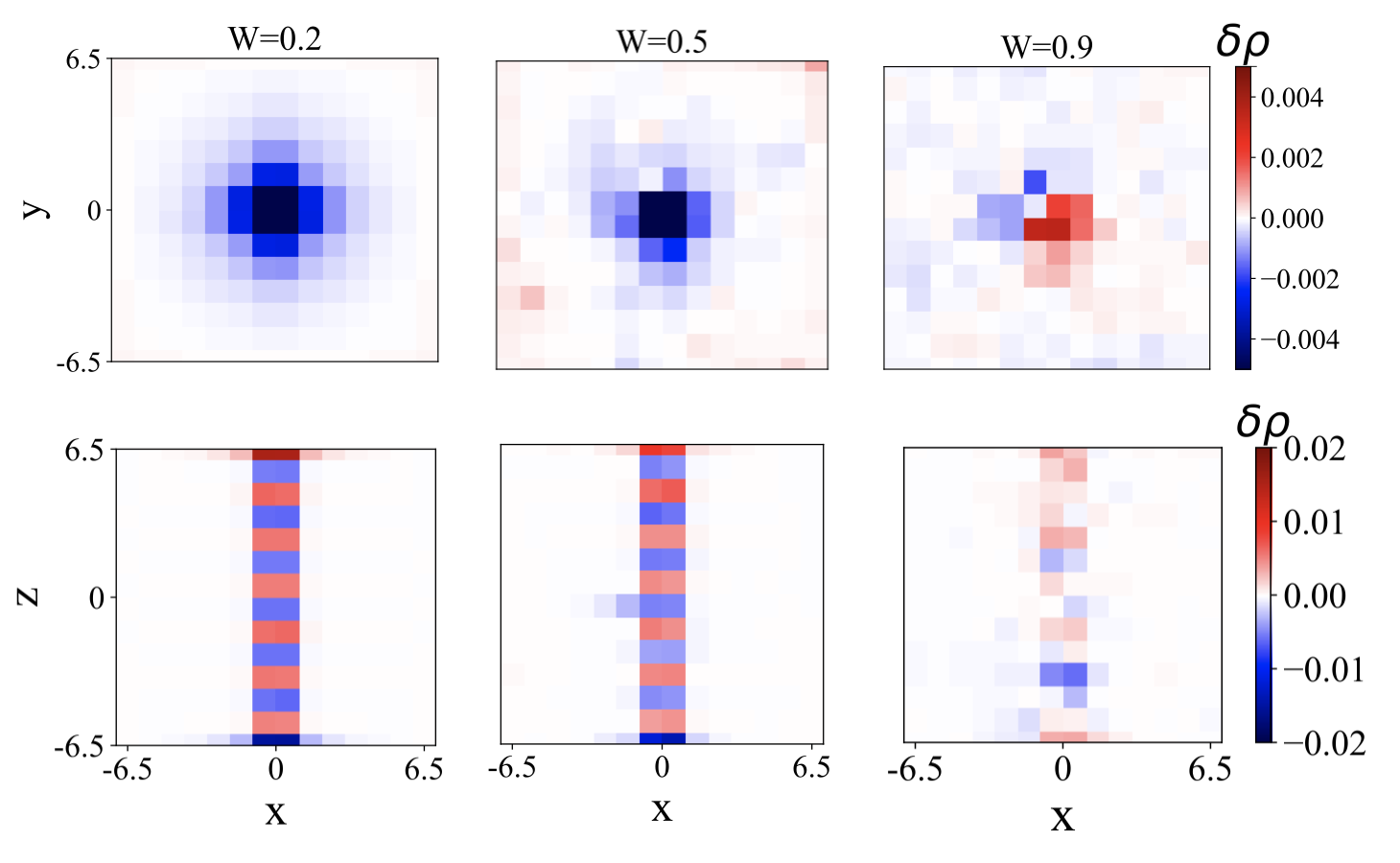}
    \caption{{\bf Topological Surface Bound Charge}: Induced charge density upon insertion of flux tube along the $z$ direction of strength $\phi/\phi_{0}=1/4$ for a system of size $L=14$ and averaging over 50 disorder configurations. Disorder strength is varied and induced charge density on (top row) $z$-surface and (bottom row) $y=0$ plane is tracked. }
    \label{fig:FluxTubeDens}
\end{figure*}

\par 
As stated in the main-body, our gauge-choice for the inserted monopole produces a flux tube which pierces the $z$ surface. Such a flux tube offers a quantitative real-space probe of the surface topology as in two-dimensions the presence of a non-zero surface Chern number, $C$, leads to the flux tube displaying bound charge $\delta Q=eC\phi/\phi_{0}$ as has been extensively studied in anomalous quantum Hall systems~\cite{Qi2011,SpinChargeVishwanath,QiSpinCharge,slager2012,MESAROS2013977}. It is therefore possible to probe the surface Chern number by measuring the charge bound to the flux tube where it pierces the $z$ surface. However, to accomplish this measurement, following Ref. \cite{wormhole}, the monopole charge must tuned away from an integer value. A simple alternative to this procedure is insertion of a flux tube rather than monopole along the $z$ direction at the origin. The insertion of a flux tube is simulated using the gauge
\begin{eqnarray}\label{eq:gauge}
A_{y}(x,y)=\phi\Theta(x)\delta(y). 
\end{eqnarray}
In examining charge bound to the flux tube, we must be aware that by tuning the flux strength can give rise to a wormhole effect for charge between the $z$ surfaces as detailed in Ref.~\cite{wormhole}. As a result, if we wish to use induced charge on the flux tube as an indicator of the surface spin-Chern number we must tune the flux strength away from $\phi=\phi_{0}/2$ as this relation for bound charge remains valid only for $\phi<\phi_{0}/2$. Following Ref.~\cite{wormhole}, we consider the case $\phi/\phi_{0}=1/4$. Modifying Eq.~\eqref{eq:deltaq} to the form,
\begin{equation}\label{eq:deltaqv}
\delta Q(R)=\sum_{\sqrt{x_{i}^2+y_{i}^2}<R, z_{i}>0}\Delta \rho(x_{i},y_{i},z_{i}),
\end{equation}
we compute induced charge contained within a cylinder extending through the upper half of the sample ($z>0$) with radius $r$ centered at the origin. The results, shown in Fig.~\ref{fig:CMLayers}(b), demonstrate that the flux tube supports a finite bound charge which is in accordance with the expected value $\delta Q=e/8$ and vanishes for increasing disorder strength, in correspondence with the analysis of the surface Chern marker. 
\par
We provide figures detailing the induced charge density averaged over 50 disorder configurations for a system of size $L=14$ in the presence of a flux tube of strength $\phi/\phi_{0}=1/4$ for $W=0.2$, $W=0.5$, and $W=0.9$ in Fig.~\ref{fig:FluxTubeDens}. These figures demonstrate clear domain walls within the bulk as a result of the staggered Zeeman field. However, we note the delocalization of charge on the surface for $W=0.5$ in correspondence with trivialization of the surface Chern number and break down of bulk domain walls for $W=0.9$.

\section{Proximity coupling of an s-wave superconductor}
In order to diagnose a transition in the bulk-boundary correspondence as a function of the disorder strength we consider proximity coupling of an s-wave superconductor. The resulting $s$-wave singlet, Bogoliubov-de Gennes (BdG) Hamiltonian takes the form, 
\begin{equation}\label{eq:BDG}
    H^{BdG}=\begin{bmatrix}
    H-\mu && \Delta \\
    \Delta^{\dagger} && -H^{*}+\mu
    \end{bmatrix},
\end{equation}
where $H$ is defined in Eq.~\eqref{eq:model} and $\Delta=\delta_{0}\tau_{0}\sigma_{2}$ where $\delta_{0}$ is the pairing strength. We make this pairing choice as, for $W<W_{c,B}$, it preserves the bulk gap as the chemical potential is varied. Furthermore, it is the utilized in Refs. \cite{Hosur2011,Rossi2020} to study the vortex phase transition. In this form, on-site potential disorder enters the Hamiltonian as a modification to the chemical potential, $\mu \rightarrow \mu + V(\mathbf{r})$. To model the vortex line we follow Ref. \cite{Rossi2020}, modifying the pairing term to the form, $\Delta(\textbf{r})=\Delta \tanh(r/\xi_{0})e^{i\phi_{0}}$, where $r=\sqrt{x^2+y^2}$, $\phi_{0}=\tan^{-1}(y/x)$, and we fix $\xi_{0}=1$. As in Refs. \cite{Hosur2011,Rossi2020}, we neglect contributions from the vector potential and Zeeman term from the field used to induce the vortex. 

\begin{figure*}
    \centering
    \includegraphics[width=16cm]{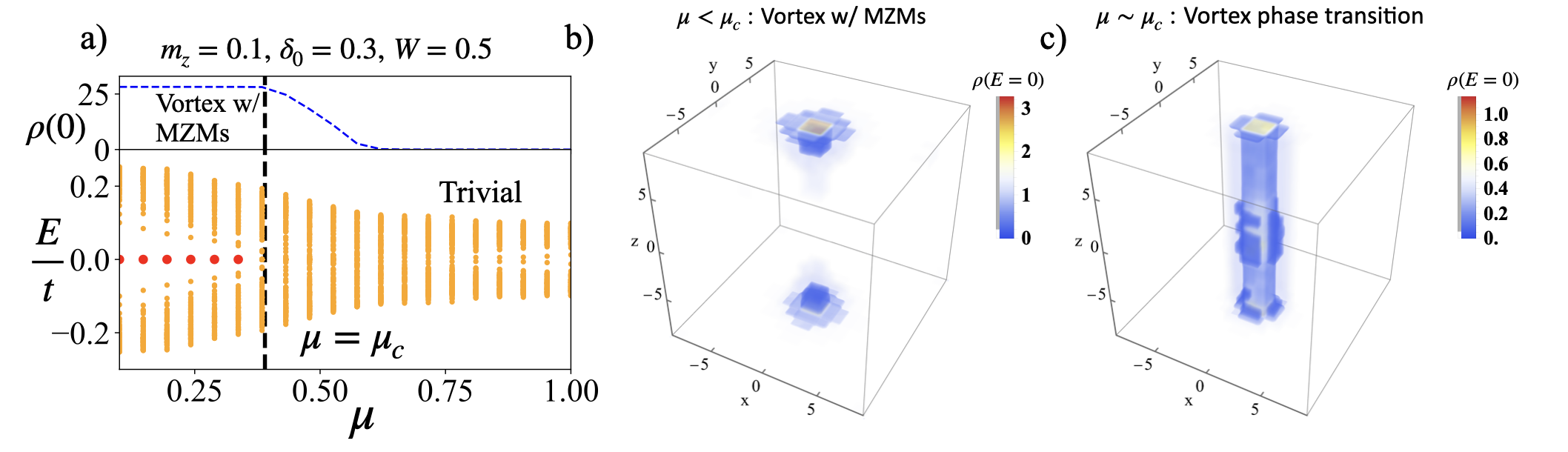}
    \caption{{\bf Vortex Phase Transition due to a Single Dirac Cone}: Spectra of Eq.~\eqref{eq:BDG}, considering the fifty eigenstates of smallest magnitude as a function of the chemical potential $\mu$, fixing $W=0.5,\; m_{z}=0.1$ and $\delta_{0}=0.3$, repeated for 50 disorder configurations. Upper panel displays disorder averaged density of states at zero energy. Above images display spatially resolved, zero energy density of states. (b)-(c) Density of states averaged for 50 disorder configurations for eq. \eqref{eq:BDG}, for the same parameters.}
    \label{fig:VPTMu}
\end{figure*}
\subsection{Vortex phase transition by tuning chemical potential}
\par
In the main body we detail how a vortex phase transition can be induced by tuning the disorder strength. In previous works\cite{Hosur2011,Rossi2020} focusing on clean systems the vortex phase transition was instead illustrated by tuning the chemical potential. Here we demonstrate that in the region $W_{c,S}<W<W_{c,B}$, such that the surface supports a gapless Dirac cone, we observe the same vortex phase transition observed in Refs. \cite{Hosur2011,Rossi2020}.
\par 
In order to estimate the critical value of the chemical potential at which the VPT occurs, $\mu_{c}$, we compute the the fifty eigenvalues of smallest magnitude using the Lanczos algorithm\cite{lanczos1950iteration} fixing $W=0.5,\; m_{z}/m=0.1$. Repeating this procedure for fifty disorder configurations yields the spectra in Fig.~\ref{fig:VPTMu}(a), demonstrating that $\mu_{c}\approx 0.4$. Examining the spatial localization of the zero energy states for $\mu\leq \mu_{c}$ in Fig.~\ref{fig:VPTMu}(b) shows that they are MZMs. The vortex phase transition is also visible for $\mu\approx \mu_{c}$ in Fig.~\ref{fig:VPTMu}(c). These results align precisely with the known results for a topological insulator with protected gapless surface states\cite{Hosur2011,Rossi2020}.

\subsection{Trivial phase zero energy density of states}
\par
\begin{figure*}
    \centering
    \includegraphics[width=16cm]{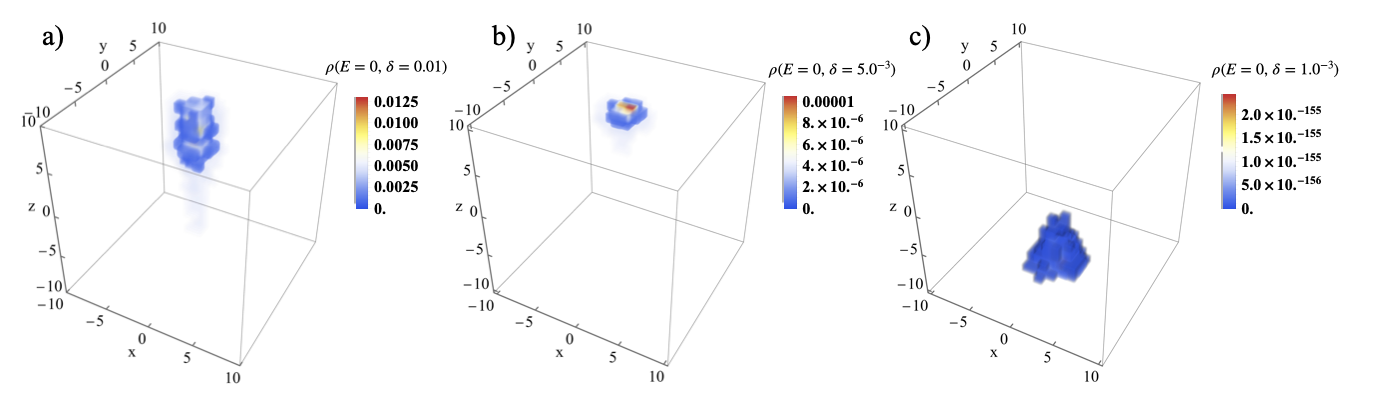}
    \caption{{\bf Trivial state accessed by vortex phase transition}: Local density of states at zero energy for eq.\eqref{eq:BDG} upon tuning the chemical potential $\mu$ such that $\mu>\mu_{c}$ and fixing $W=0.5$, $\delta_{0}=0.3$ and $m_{z}=0.1$ for a system size $L=20$ and averaging over 50 disorder configurations. The broadening parameter used in computing the local density of states, $\delta$ is set to (a) $\delta=0.01$, (b) $\delta=0.005$ and (c) $\delta=0.001$ demonstrating a vanishing local density of states as the broadening is decreased.}
    \label{fig:FluxTubeDensTriv}
\end{figure*}

 
In the previous section and the main body, the presence of Majorana zero modes and a vortex phase transition via tuning the chemical potential ($\mu$) or disorder strength ($W$) of the BdG Hamiltonian was demonstrated. Evidence for the Majorana zero modes  and the vortex phase transition were provided. Here we demonstrate that if we tune the chemical potential into the trivial regime ($\mu > \mu_{c}$) zero energy states are absent. This is shown in Fig.~\ref{fig:FluxTubeDensTriv}, where the disorder averaged local density of states is plotted as a function of the broadening parameter, $\delta$, used in computation of the local density of states. We note that as the broadening parameter is decreased the local density of states correspondingly vanishes indicating a lack of zero energy states. 
\section{Comparison to two-dimensional Chern insulator within GUE}
\par 
One of the primary discoveries of this work is the existence of a range of disorder strengths and Zeeman field strengths over which the bulk mobility remains intact while the surface supports a two-dimensional GUE metal. Importantly, throughout this parameter range the surface states are extended.
\par 
This is a novel finding as in familiar two-dimensional Chern insulators, a gapless Dirac cone is identified only at the critical point between topological and trivial insulator\cite{prodan2011disordered,prodanets,Prodan2009,Prodan2013,moreno2023topological}. The gapless Dirac cone \emph{does not} exist over a range of parameters. Rather all states at the Fermi energy are localized on either side of the critical point as a function of disorder strength. This is conveyed schematically in Fig. 1 of the main body. 
\par
\begin{figure*}
    \centering
    \includegraphics[width=15cm]{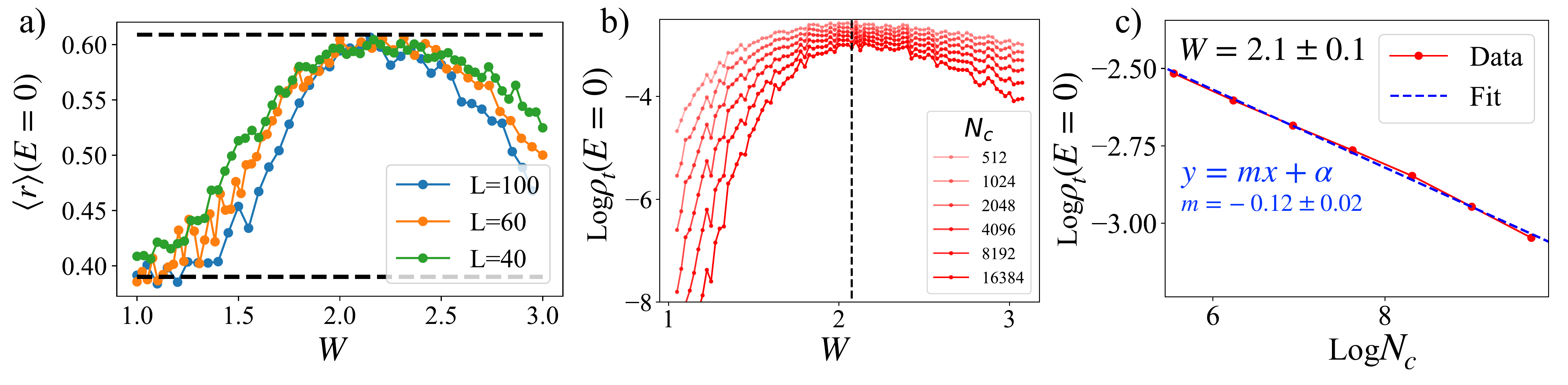}
    \caption{{\bf Disordered two-dimensional Chern insulator:} (a) Adjacent gap ratio as defined in Eq.~\eqref{eq:LevelSpace} for $E=0$ as a function of disorder strength, $W$, averaging over 500 disorder configurations for three system sizes. Away from the critical point the results align with the expected result for a Poisson distribution, $\langle r \rangle = 2\log 2-1 \approx 0.39$. At the critical point the adjacent gap ratio approaches the expected result for a GUE, $\langle r \rangle \approx 0.602$. (b) Typical density of states at $E=0$ varying the KPM expansion order $N_{c}$ for a system of size $L=2000$ and averaging over 5000 disorder configurations. (c) Zero energy typical density of states at the critical point as a function of the KPM expansion parameter. Data is shown on a Log-Log scale with the results of a power law fit following Eq. \eqref{eq:scaling}. }
    \label{fig:2DChern}
\end{figure*}

 
Here we provide the results of computing typical density of states and level statistics for a two-dimensional Chern insulator to illustrate the distinct behavior to that observed on the surface of an axion insulator. The model we study takes form, 
\begin{equation}\label{eq:chernmodel}
    H=m\sum_{l}c^{\dagger}_{l}\tau^{z}c_{l}+\frac{t}{2}\sum_{ll'}'c^{\dagger}_{l}\tau^{z}c_{l'}+\frac{-it}{2}\sum_{ll'}'c^{\dagger}_{l}\hat{\mathbf{n}}_{ll'}\cdot \mathbf{\tau}c_{l'} +\sum_{l}V(l)c^{\dagger}_{l}c_{l},
\end{equation}
where nearest neighbor hoppings are denoted by sums over $\langle l, l' \rangle$, $\hat{\mathbf{n}}_{ll'}$ is a two-component nearest-neighbor unit vector and we fix, $m/t=1.75$ such that the Chern number is of unit strength. First we compute the level statistics fixing $L=40,60,100$, averaging over 500 disorder configurations with the results shown in Fig.~\eqref{fig:2DChern}(a). It is clear from this figure that away from the critical point, $W_{c}\approx 2.1$, all zero energy states are localized with the average level spacing approaching the expected value for a Poisson distribution for increasing system size. However, at the critical point the average level spacing appears constant at the expected value for a GUE. This is in stark contrast to Fig.~\eqref{fig:levelstats}(a) for which the extended states at zero energy withing the GUE persist upon closure of the surface mobility gap. 
\par 
We further compute TDOS at zero energy for the same model fixing $L=2000$ and averaging over 5000 disorder configurations. The results as a function of the expansion paramter $N_{c}$, are shown in Fig.~\eqref{fig:2DChern}(b). In this figure it is clear that, away from the critical point, the TDOS converges rapidly to zero with increasing $N_{c}$. Near the critical point, $W= 2.1 \pm 0.1$, 
the TDOS continues to decrease with increasing $N_{c}$. This is expected following the scaling form at the band center ($E=0$) given in the main body, 
\begin{equation}\label{eq:scaling}
   \rho_{t}(E=0)\sim N_{c}^{-\beta/(z\nu)} f\left((W-W_{c})N_C^{1/(\nu z)}\right).
\end{equation}
Fixing $z=d=2$ and utilizing the known values of $
\beta$, and $\nu$ for a two-dimensional Chern insulator, we expect $\rho_{t}(E=0)\sim N_{c}^{-\beta/(z\nu)}\approx N_{c}^{-0.12}$ at the critical location\cite{slevin1999,Brndiar2006,slevin2009,slevin2011,janssen1998statistics,ujfalusi2015}. In Fig.~\eqref{fig:2DChern}(c) we fit the TDOS at the critical point as a function of $N_{c}$. Fitting the data reveals that $\rho_{t}(E=0)\sim N_{c}^{-x}$, where $x\approx 0.12 \pm 0.02$. 
This is in excellent agreement with the expected value. These results underscore the novelty of the behavior of the Chern insulator on the surface of an axion insulator explored in the main body.
\bibliography{ref.bib}